\documentclass[JCAP]{article} 
\usepackage{jcappub}

\usepackage{epstopdf}
\usepackage{graphicx}
\usepackage{dcolumn}
\usepackage{bm}
\usepackage{color}
\usepackage{hyperref}
\usepackage{multirow}
\input epsf

\newcommand \Mnu{$\Sigma_i m_{\nu_i}$ }

\begin{document}
\title{Cosmology with massive neutrinos I: towards a realistic modeling of the relation between matter, haloes and galaxies}

\author[a]{Francisco Villaescusa-Navarro,} \author[b,c,d]{Federico Marulli,} 
\author[a,e]{Matteo Viel,} \author[f,g,h]{Enzo Branchini,}
\author[i]{Emanuele Castorina,} \author[j,k]{Emiliano Sefusatti,}
\author[h]{Shun Saito}  

\affiliation[a]{INAF - Osservatorio Astronomico di Trieste, Via Tiepolo 11, 34143, Trieste, Italy}
\affiliation[b]{Dipartimento di Fisica e Astronomia - Universit\`a di Bologna, 
viale Berti Pichat 6/2, I-40127 Bologna, Italy}
\affiliation[c]{INAF - Osservatorio Astronomico di Bologna, via Ranzani 1, I-40127 Bologna, Italy}
\affiliation[d]{INFN - Sezione di Bologna, viale Berti Pichat 6/2, I-40127 Bologna, Italy}
\affiliation[e]{INFN sez. Trieste, Via Valerio 2, 34127 Trieste, Italy}
\affiliation[f]{Dipartimento di Matematica e Fisica, Universit\`a degli Studi Roma Tre, via della Vasca Navale 84, 00146 Roma, Italy}
\affiliation[g]{INFN - Sezione di Roma Tre, via della Vasca Navale 84, I-00146 Roma, Italy}
\affiliation[h]{INAF - Osservatorio Astronomico di Roma, via Frascati 33, I-00040 Monte Porzio Catone (RM), Italy}
\affiliation[i]{SISSA - International School For Advanced Studies, Via Bonomea, 265 34136 Trieste, Italy}
\affiliation[j]{The Abdus Salam International Center for Theoretical Physics, Strada Costiera 11, 34151, Trieste, Italy}
\affiliation[k]{INAF, Osservatorio Astronomico di Brera, Via Bianchi 46, I-23807 Merate (LC)  Italy}
\affiliation[h]{Kavli Institute for the Physics and Mathematics of the Universe (WPI), Todai Institutes for Advanced Study, The University of Tokyo, Chiba 277-8582, Japan}

\emailAdd{villaescusa@oats.inaf.it}

\abstract{By using a suite of large box-size N-body simulations that
  incorporate massive neutrinos as an extra set of particles, 
  with total masses of 0.15, 0.30, and 0.60 eV, we investigate the 
  impact of neutrino masses on the spatial distribution of dark matter haloes and on 
  the distribution of galaxies within the haloes. We compute the bias
  between the spatial distribution of dark matter haloes and the overall
  matter and cold dark matter distributions using statistical tools such
  as the power spectrum and the two-point correlation function. 
  Overall we find a scale-dependent bias on large
  scales for the cosmologies with massive neutrinos. In particular,
  we find that the bias decreases with the scale, being this effect more
  important for higher neutrino masses and at high redshift. 
  However, our results indicate that the scale-dependence in the bias is reduced
  if the latter is computed with respect to the cold dark matter distribution only. 
  We find that the value of the bias on large scales is reasonably well 
  reproduced by the Tinker fitting formula once the linear cold dark 
  matter power spectrum is used, instead of the total matter power 
  spectrum. We also investigate whether scale-dependent bias really 
  comes from purely neutrino's effect or from nonlinear gravitational collapse 
  of haloes. For this purpose, we address the $\Omega_\nu$-$\sigma_8$ degeneracy
  and find that such degeneracy is not perfect, implying that neutrinos 
  imprint a slight scale dependence on the large-scale bias.
  Finally, by using a simple halo occupation distribution (HOD) model,
  we investigate the impact 
  of massive neutrinos on the distribution of galaxies within dark matter
  haloes. We use the main galaxy sample in the Sloan Digital Sky Survey 
  (SDSS) II Data Release 7 
  to investigate if the small-scale galaxy clustering alone can be used to
  discriminate among different cosmological models with different neutrino masses. 
  Our results suggest that different choices of the HOD parameters can 
  reproduce the observational measurements relatively well, and we 
  quantify the difference between the values of the HOD parameters 
  between massless and massive neutrino cosmologies.}
  
\maketitle

\section{Introduction}

Neutrinos are one of the most interesting and enigmatic particles of
the particle standard model. Postulated by Wolfgang Pauli in 1930 
to avoid the violation of energy, momentum and spin in the
beta decay process, neutrinos were eventually detected in 1956 by Cowan and
Reines \cite{CowanReines}. Neutrinos have long been considered massless 
particles until the "oscillation" phenomenon, i.e. the change of flavor, was 
detected in the neutrinos produced within the sun \cite{Cleveland}.
The latest results that combine solar, atmospheric and reactor neutrinos allow to constrain 
the (squared) mass difference between the three neutrino mass eigenstates:  
$\bigtriangleup m^2_{12}=7.5\times10^{-5}$ ${\rm eV}^2$ and $|\bigtriangleup
m^2_{23}|=2.3\times10^{-3}$ ${\rm eV}^2$ \cite{Fogli,Tortola}. From the 
theoretical side, it is of great interest to determine the absolute neutrino mass scale, since 
this may reveal physics beyond the particle standard model.

From a cosmological point of view it is mandatory to account for the neutrino
masses in a precision cosmology era as the one we have just
entered. The Big Bang theory predicts the existence of a cosmic
neutrino background, and those neutrinos play a fundamental role in
setting the primordial abundance of light elements. Moreover, their
masses impact on cosmology, at the linear level, on two different
ways: modifying the matter-radiation equality and slowing down the
growth of matter perturbations. The combination of these effects
results in a suppression of the matter power spectra on small scales
\cite{LesgourguesPastor}. The shape of the matter power spectra,
constrained on very large scales by the cosmic microwave background
and by the galaxy distribution on intermediate scales, has been used
to put upper limits on the neutrino masses. Present limits, obtained
independently from different galaxy surveys such as 
SDSS\footnote{http://www.sdss.org/} and
CFHTLS\footnote{http://www.cfht.hawaii.edu/Science/CFHLS/} 
point out that \Mnu$<0.3$ eV at $2\sigma$
\cite{Hannestad_2003, Reid, Thomas, Swanson, Saito_2010, 
dePutter, Xia2012, WiggleZ, Zhao2012, 2013MNRAS.430..747M}, 
although by using data from the WiggleZ\footnote{http://wigglez.swin.edu.au/site/}
survey it is found that \Mnu $< 0.15$ eV ($95\%$) \cite{Riemer-Sorensen},
and it is expected that 
those limits shrink to \Mnu$\lesssim0.03$ eV, i.e. allowing a 
determination of the neutrino masses with a significance 
of $2\sigma$ even in the normal
 hierarchy, with future galaxy surveys as Euclid 
 \cite{Costanzi,Basse,carbone2011}. It is worth to
note that the latest results from Planck \cite{Planck_2013} gives an
upper limit on the sum of the neutrino masses equal to \Mnu $< 0.93$
eV within a $95\%$\footnote{This is found with
  Planck alone, i.e. without combining results with BAO,
  high-$l$...etc.} confident interval, assuming a flat 
  $\Lambda {\rm CDM}$ cosmological model.

Although the impact of neutrino masses on cosmology is very well
understood at the linear order \cite{LesgourguesPastor,
LesgourguesBook}, their impact on the fully non-linear regime 
has not been extensively studied. 
The non-linear regime is important at both low redshift and
small scales, and N-body techniques emerged as the best tools to 
follow structure formation in it. However, whereas the cold dark matter 
(CDM), baryons, stars  and black holes are commonly simulated in 
N-body simulations, cosmological neutrinos have not received the same attention. 

Recent works have investigated the impact of massive neutrinos
on the non-linear regime by using either semi-analytic models or
N-body simulations. In particular, those works considered 
the clustering of relic neutrinos within the gravitational potential wells of CDM 
haloes \cite{Ma, Wong, Brandbyge_haloes, Villaescusa-Navarro_2011,
Villaescusa-Navarro_2013a} and the impact of neutrino masses on: the
non-linear matter power spectrum \cite{Brandbyge2008, Saito_2008,
Brandbyge2009, BrandbygeHybrid, Saito_2009, Viel_2010, 
Agarwal2011, Bird_2011, Wagner2012}, the
Lyman-$\alpha$ forest \cite{Viel_2010, Villaescusa-Navarro_2012}, the
halo mass function \cite{Brandbyge_haloes, Marulli_2011,
Villaescusa-Navarro_2012} and the redshift-space distortions
\cite{Marulli_2011}. Whereas most of those works are focused on the impact
of neutrino masses on large scales, in this paper we explore the
effects of massive neutrinos on scales comparable with those of galaxies and
dark matter haloes. To achieve that, we have run a large set
of N-body simulations containing CDM and neutrinos particles,
simulating many different cosmological models characterized by different  
sums of  neutrino masses. Our simulations explore
different total neutrino masses (assuming 3 degenerate neutrino
families): 0.00, 0.15, 0.30 and 0.60 eV. Although observational constraints 
prefer a value of \Mnu smaller than 0.30 eV, a value of 0.60 eV is not
completely unreasonable, as the latest results of Planck indicate. Moreover,
our aim in this paper is not limited to simulate realistic cosmological models,
but, more in general, to investigate how massive neutrinos affect the spatial 
distribution of dark matter haloes and the clustering properties of the galaxies
residing in those.

The clustering properties of dark matter
haloes in cosmologies with massive neutrinos have already been
studied in \cite{Marulli_2011}. In that paper, authors used N-body
simulations incorporating neutrinos through the so-called \textit{grid
  method} (see \cite{Brandbyge2009,Viel_2010}). In this method
neutrinos only contribute to the long distance force through the 
particle mesh (PM) method, providing a
fast implementation of the matter evolution in neutrino cosmologies,
but with the limitation that it does not properly capture the non-linear
neutrino regime. The use of the grid method is, thus, only justified
on regimes where the non-linear neutrino effects are negligible (at $z\gtrsim2$ and on
large linear scales \cite{Brandbyge2009, Viel_2010, Villaescusa-Navarro_2013a}).  

In this paper we study the impact of neutrino masses on the halo-matter and
galaxy-matter bias by using N-body simulations that incorporates neutrinos as particles 
(using the so-called \textit{particle method}). In comparison with the \textit{grid}-based neutrino
N-body simulations, these simulations are better suited to follow the 
fully non-linear regime. We investigate the halo-matter 
bias using statistical tools such as the power spectrum and the two-point correlation 
function. We then populate with realistic galaxies the dark matter haloes of the different 
cosmological models using a simple halo occupation distribution (HOD) model and 
investigate the galaxy clustering properties in universes with massive neutrinos.
We notice that baryonic effects can also impact on the clustering properties of dark 
matter haloes and galaxies (see for instance \cite{2013arXiv1310.7571V}).

This paper is the first of a series of three papers \cite{Castorina,Costanzi2}. In 
Paper II \cite{Castorina} we investigate the universality of the halo mass function (HMF)
and of halo bias in cosmologies with massive neutrinos. We show that the abundance of Friends-of-Friends haloes in a massive neutrino model is well reproduced by the Crocce et al. (2010) \cite{Crocce_2010} fitting formula once the
CDM linear power spectrum is used to compute $\sigma(M,z)$, rather than the total matter 
power spectrum. If the above prescription is adopted, then the HMF becomes universal with respect to neutrino masses.
Paper II also presents similar results for the large-scale halo bias in presence of massive neutrinos. In particular we show that universality of linear bias factors is recovered only if they are computed with respect to the spatial distribution of the CDM component alone.

In Paper III \cite{Costanzi} we verify the findings of Paper II by showing that the HMF of
Spherical Overdensity haloes is well described by the Tinker et al. (2008) \cite{Tinker_2008}
once the CDM linear matter power spectrum is used to calculate $\sigma(M,z)$. 
We investigate the effect that the new prescription
for the HMF has on cosmological parameters inferred from cluster abundance data, using 
as a case study the Planck SZ-selected sample of clusters. We find that, for a cosmology with massive
neutrinos, the improved HMF calibration provides a stronger degeneracy between the
cosmological parameters $\sigma_8$ and
$\Omega_m$, which leads to a lower $\sigma_8$ mean value. 
Taking into account such an
effect has the consequences of increasing the tension between the cosmological parameters derived
from Planck CMB data and those from cluster number counts~\cite{2012arXiv1212.6267H}.

The paper is organized as follows. We first describe in section
\ref{Simulations} the N-body simulations we have carried out for this
work. In section \ref{haloes_clustering} we investigate the impact of neutrino
masses on the spatial distribution of dark matter haloes by means of  
the power spectrum and the correlation function. Then, in section \ref{sec:HOD},
we populate the dark matter haloes 
of the different cosmological models with galaxies,
studying how the galaxy clustering properties are affected by the presence
of massive neutrinos. Finally, we present the main conclusions of this work 
in section \ref{Conclusions}.

\section{The simulations}
\label{Simulations}

We have run N-body simulations containing CDM and neutrino particles, 
using the particle-based method, to study the effects of 
massive neutrinos on the spatial distribution of dark
matter haloes and galaxies. As discussed in the above section, the simulations
run with this method are better suited for this work, since they allow us to capture
the fully non-linear regime, as opposed to the simulations used in \cite{Marulli_2011}.
On the other hand, the disadvantage is that the 
computational cost of running the simulations with this method 
is much higher than with the grid-based method. 

The simulations have been run using the
TreePM code GADGET-3, an improved version of the code GADGET-2
\cite{Springel_2005}, using the particle-based
implementation. In this implementation, neutrinos are treated as an
extra set of particles, in the same way as the CDM, with the
difference that, at the starting redshift of the simulation, the
neutrinos receive an extra thermal velocity component obtained by
random sampling the neutrino Fermi-Dirac linear\footnote{
In \cite{Villaescusa-Navarro_2013a} it was shown that this distribution
is impacted by non-linear effects at low redshift.} momentum
distribution. On small scales, the force affecting the neutrinos is
computed using the short-range tree. As pointed out in
\cite{Villaescusa-Navarro_2013a}, this feature is required to
correctly account for the clustering of neutrinos within dark matter
haloes and to reproduce the neutrino haloes down to small scales. For
further details about this method we refer
the reader to \cite{Brandbyge2008, Viel_2010, Bird_2011}. The code
time-step is set by the CDM particles, independently on the presence
of neutrino particles. The typical total CPU time consumption of our N-body
simulations is between 1,000 and 10,000 hours. 

The initial conditions of the N-body simulations have been generated
at $z=99$, using the Zel'dovich approximation for both the CDM and the
neutrino particles. The transfer functions have been obtained through CAMB
\cite{CAMB}. We have incorporated the baryon effects 
(for instance the BAO peaks) into the CDM particles
by using a transfer function that
is a weighted average of the transfer functions of the CDM and the
baryons, the formers as given by CAMB. The gravitational softening of
both particle types have been set to $1/30$ of their mean
inter-particle linear spacing.

\begin{table}
\begin{center}
\resizebox{15.5cm}{!}{
\begin{tabular}{|c|c|c|c|c|c|c|c|c|c|c|c|}

\hline
Name & \Mnu & Box  & $\Omega_{\rm m}$ & $\Omega_{\rm b}$ & $\Omega_\nu$ & $\Omega_\Lambda$ & $h$ & $n_s$ & $N_\mathrm{CDM}^{1/3}$ & $N_\nu^{1/3}$ & $\sigma_8$ \\
 & (eV) & ($h^{-1}\rm{Mpc}$) & & & & & & & & & (z=0)\\
\hline
\hline 
H0-Planck-LV (1) & $0.00$ & $1000$ & 0.3175 & 0.049 & 0.0 & 0.6825 & 0.6711 & 0.9624 & $1024$ & $0$ & $0.834$\\
\hline
L6-Planck (1) & $0.60$ & $500$ & 0.3175 & 0.049 & 0.0143 & 0.6825 & 0.6711 & 0.9624 & $512$ & $512$ & $0.690$\\
\hline
L3-Planck (1) & $0.30$ & $500$ & 0.3175 & 0.049 & 0.0072 & 0.6825 & 0.6711 & 0.9624 & $512$ & $512$ & $0.761$\\
\hline
L15-Planck (1) & $0.15$ & $500$ & 0.3175 & 0.049 & 0.0036 & 0.6825 & 0.6711 & 0.9624 & $512$ & $512$ & $0.799$\\
\hline
L0-Planck (1) & $0.00$ & $500$ & 0.3175 & 0.049 & 0.0 & 0.6825 & 0.6711 & 0.9624 & $512$ & $0$ & $0.834$\\
\hline \hline

H6 (8) & $0.60$ & $1000$ & 0.2708 & 0.050 & 0.0131 & 0.7292 & 0.7 & 1.0 & $512$ & $512$ & $0.675$\\
\hline
H3 (8) & $0.30$ & $1000$ & 0.2708 & 0.050 & 0.0066 & 0.7292 & 0.7 & 1.0 & $512$ & $512$ & $0.752$\\
\hline
H0 (8) & $0.00$ & $1000$ & 0.2708 & 0.050 & 0.0 & 0.7292 & 0.7 & 1.0 & $512$ & $0$ & $0.832$\\
\hline
H0-HR (1) & $0.00$ & $1000$ & 0.2708 & 0.050 & 0.0 & 0.7292 & 0.7 & 1.0 & $768$ & $0$ & $0.832$\\
\hline
H6-LR (20)& $0.60$ & $1000$ & 0.2708 & 0.050 & 0.0131 & 0.7292 & 0.7 & 1.0 & $256$ & $256$ & $0.675$\\
\hline
L6 (1) & $0.60$ & $500$ & 0.2708 & 0.050 & 0.0131 & 0.7292 & 0.7 & 1.0 & $512$ & $512$ & $0.675$\\
\hline
L3 (1) & $0.30$ & $500$ & 0.2708 & 0.050 & 0.0066 & 0.7292 & 0.7 & 1.0 & $512$ & $512$ & $0.752$\\
\hline
L0 (1) & $0.00$ & $500$ & 0.2708 & 0.050 & 0.0 & 0.7292 & 0.7 & 1.0 & $512$ & $0$ & $0.832$\\
\hline
L0-HR (1) & $0.00$ & $500$ & 0.2708 & 0.050 & 0.0 & 0.7292 & 0.7 & 1.0 & $768$ & $0$ & $0.832$\\
\hline\hline

H6s8 (8) & $0.06$ & $1000$ & 0.2708 & 0.050 & 0.0131 & 0.7292 & 0.7 & 1.0 & $512$ & $512$ & $0.832$\\
\hline
H0s8 (8) & $0.00$ & $1000$ & 0.2708 & 0.050 & 0.0 & 0.7292 & 0.7 & 1.0 & $512$ & $0$ & $0.675$\\
\hline
H0s8-CDM (8) & $0.00$ & $1000$ & 0.2708 & 0.050 & 0.0 & 0.7292 & 0.7 & 1.0 & $512$ & $0$ & $0.701$\\
\hline
L6-1 (1) & $0.60$ & $500$ & 0.2708 & 0.050 & 0.0131 & 0.7292 & 0.7 & 1.0 & $512$ & $512$ & $0.832$\\
\hline\hline

L6-2 (1) & $0.60$ & $500$ & 0.3000 & 0.050 & 0.0131 & 0.7000 & 0.7 & 1.0 & $512$ & $512$ & $0.749$\\
\hline
\end{tabular}
}
\end{center} 
\caption{Summary of the simulations used in the present
  work. The simulations have been divided into four different 
  groups (see text for details). The sizes of the simulations
  boxes are either 500 or 1000 $h^{-1}$ Mpc. The name of the
  simulations with box sizes of 1000 $h^{-1}$ Mpc starts with H (from Huge),
  whereas the name of simulations with box sizes of 500 $h^{-1}$ Mpc begins
  with L (from Large).
  $m_{\nu_i}$ is the mass of a single neutrino species, and \Mnu
  is the sum of the neutrino masses. The number of
  independent simulations for each cosmological model is shown in
  parentheses after the name of each simulation.}
\label{tab_sims}
\end{table}
 
In table \ref{tab_sims} we list the different simulations we have run
for this paper along with the values of their cosmological
parameters. The size of our simulation boxes are either 500 or 1000
$h^{-1}$Mpc, chosen to have large statistics but keeping moderate the
computational cost. We have used the simulations with large box sizes 
(1000 $h^{-1}$Mpc) to investigate the spatial distribution of dark matter
haloes on large scales, whereas we have used the simulations 
with 500 $h^{-1}$Mpc to explore the galaxy clustering properties. 
The number of CDM particles in the simulations is
$512^3$, except for a few cases that we have run to perform
convergence tests. The masses of the CDM particles for the simulation
with the highest resolution, L0-HR, are of $2.07\times10^{10}~h^{-1}{\rm
  M}_\odot$.  We have simulated massive neutrino cosmologies with a
sum of the neutrino masses equal to 0.15, 0.30 and 0.60 eV, assuming
that the three neutrino masses are degenerate, to investigate the dependence
of our results on the neutrino masses. In simulations
containing neutrino particles, the number of those is $512^3$, except
for the H6-LR series that were run to study the errors when computing
the different power spectra and bias. In
\cite{Villaescusa-Navarro_2013a} it was shown that this number of
neutrino particles is enough to properly capture the neutrino effects
on the large scale structure. For all the simulations we have saved
snapshots at $z=0$, 0.5, 1 and 2.

All the simulations corresponds to flat, $\Omega_\Lambda=1-\Omega_{\rm m}$,
cosmological models.
Our simulations can be organized into 4 different groups as shown in the table
\ref{tab_sims}. The simulations belonging to the group 1 (H0-Planck-LV,
L6-Planck, L3-Planck, L15-Planck and L0-Planck) share the values of the cosmological parameters 
$\Omega_{\rm b}$, $\Omega_\Lambda$, $h$ and $n_s$. Moreover, the value 
of the parameters $\Omega_{\rm m}=\Omega_{\rm cdm}+\Omega_{\rm b}+\Omega_\nu$ and 
$A_s$ (the amplitude of the primordial power spectrum)
are the same for all of them. The differences among the simulations within this group
are in the value of their parameters $\Omega_{\rm cdm}$, $\Omega_\nu$ and $\sigma_8$. Since 
$\Omega_{\rm m}$ is kept constant for all the simulations within this group, simulations with higher
values of $\Omega_\nu$ have smaller values of $\Omega_{\rm cdm}$, such as their sum is 
the same for all the models. The value of the parameter $\sigma_8$ depends on the neutrino
masses since the amplitude of the power spectrum is fixed on large scales, thus, models with higher
neutrino masses will have a lower value of $\sigma_8$. We note that within this group, the model with 
massless neutrinos has cosmological values in agreement with the latest results of Planck 
\cite{Planck_2013}. To distinguish the simulations within this group we add the suffix "Planck" to their
names. The simulations within this group have been mainly used to study the clustering of galaxies.

The simulations belonging to the second group are: H6, H3, H0, H0-HR, H6-LR, L6, L3, L0, L0-HR. The
characteristic of these simulations are the same as those of group 1, i.e. they share 
the value of the parameters $\Omega_{\rm b}$, $\Omega_\Lambda$, $h$, $n_s$, 
$\Omega_{\rm m}=\Omega_{\rm cdm}+\Omega_{\rm b}+\Omega_\nu$ and $A_s$ (different from group 1) and differ in their values of 
$\Omega_{\rm cdm}$, $\Omega_\nu$ and $\sigma_8$. Whereas the cosmological parameters
of the simulation with massless neutrinos in the group 1 are in agreement with Planck, the massless 
cosmological model of group 2 is instead close to the latest results of WMAP. We have
used the simulations in this group to study both the clustering of dark matter haloes and galaxies.

Simulations in group 3 share the cosmology of group 2 except for the amplitude of initial fluctuations, $A_s$. We have used these simulations to study the $\Omega_\nu-\sigma_8$ degeneracy,  choosing $A_s$ to obtain models characterized by different neutrino masses and by the same value of $\sigma_8$. The simulations belonging to this group are: H6s8, H0s8, H0s8-CDM and L6-1, with the latter employed as well to study the impact of $\sigma_8$ on galaxy clustering properties.

Finally group 4 is constituted by the simulation L6-2 only. 
Although most of the values of the cosmological 
parameters of this simulation are equal to those of simulations in 
groups 2 and 3, this simulation has a 
higher value of $\Omega_{\rm m}$. The value of the parameter $A_s$ 
is however the same as the one of the
simulations in group 2. We have used this simulation to study the impact 
of $\Omega_{\rm m}$ on galaxy clustering.

For each simulation we identify the dark matter haloes using two different
algorithms: the Friends-of-Friends (FoF) group finder \cite{FoF} and
the SUBFIND algorithm \cite{Subfind}. We have run the FoF algorithm
with two different values of the linking length parameter: $b=0.16$ and
$b=0.2$. Both algorithms have been run only on top of the CDM component.
We have done this to avoid the contamination in mass arising from unbounded 
neutrinos. This effect can significantly bias the mass of low mass haloes. 
In the appendix \ref{Appendix_C} we show that our
results, in terms of the halo-matter bias are not affected if the haloes are instead identified
by using the total matter distribution. Moreover, in Paper II and Paper III we checked that
running the halo finder algorithms just on top of the CDM distribution has a negligible 
impact on the halo mass function. We emphasize that this effect is expected since
studies related with the neutrino clustering 
\cite{Wong,Brandbyge_haloes,Villaescusa-Navarro_2011,Villaescusa-Navarro_2013a},
have shown that the neutrino contribution to the total mass of
the halo is below $0.5\%$ for reasonable neutrino masses, as the ones
considered here.

The dark matter haloes we have used for this paper correspond
to the groups identified by SUBFIND. Those groups correspond to spherical
overdensity (SO) haloes with the virial radius taken
as the radius at which the mean interior density is 200 times that of
the universe mean density at that redshift. We will refer to the masses
of these haloes as $M_{200}$. Both algorithms, FoF and SUBFIND, provide us with 
the position of the center of the haloes, that we use when 
measuring the power spectrum or correlation function of the dark matter haloes.

\section{Clustering of dark matter haloes}
\label{haloes_clustering}

In this section we investigate the impact of massive neutrinos on the spatial 
distribution of dark matter haloes. We characterize the clustering of
dark matter haloes by using 
2-points statistics both in Fourier and in configuration space.
We choose to consider both estimators, the power spectrum and the 
2-pt correlation function since they probe different scales and have 
different sensitivity to non-linear effects and scale-dependent bias.
We start by presenting the results in terms of the power spectrum in subsection
\ref{subset:power_spectrum}, while the halo clustering in terms of the
correlation function is shown in subsection \ref{subsec:correlation_function}.
In this paper we limit our study to SO haloes, leaving the results in terms of FoF
haloes to our companion paper \cite{Castorina}.

\subsection{Halo bias: power spectrum}
\label{subset:power_spectrum}

It is well known that dark matter haloes, like galaxies, are biased
tracers of the underlying matter density field. In this section we
investigate the clustering properties of dark matter haloes in several
cosmological models with different neutrino masses.

The bias between the spatial distribution of haloes and that of matter
is usually computed as the ratio between the power spectrum of haloes
($P_{\rm hh}$) to that of the matter ($P_{\rm mm}$), that is $b^2_{\rm
hh}(k)=P_{\rm hh}(k)/P_{\rm mm}(k)$. However, this estimate is 
prone to stochasticity in the bias relation. For this reason we have 
also used the alternative estimator, $b_{\rm
hm}(k)=P_{\rm hm}(k)/P_{\rm mm}(k)$, where $P_{\rm hm}(k)$ is the
halo-matter cross-power spectrum, which, being based on 
the cross power, is less sensitive to stochasticity \cite{Dekel, 
Hamaus_2010, Baldauf_2013, Smith_2006, Baldauf_2009}.

The power spectrum is computed in the following way: we assign the
positions of the particles or halo centers to a regular cubic grid with $N\times N\times N$ points using
the cloud-in-cell (CIC) interpolation scheme. Then, we compute the value of
the density contrast, $\delta(\vec{r})=\rho(\vec{r})/\bar{\rho}-1$, at
each point of the grid. By using the Fast Fourier Transform algorithm, 
we calculate the density contrast in
Fourier space, $\delta(\vec{k})$. We correct for the CIC mass assignment scheme as 
\cite{Jing_2005, Montesano_2010}:
\begin{equation}
\delta(\vec{k})\rightarrow\frac{\delta(\vec{k})}{W^2(k)}~,
\end{equation}
where 
\begin{equation}
W(k)=\left[\prod_{i=1}^3\frac{\sin (\pi k_i/2k_N)}{(\pi k_i/2k_N)}\right]^2~,
\end{equation}
with $k=\sqrt{k_1^2+k_2^2+k_3^2}$ and $k_N=\pi N/L$ being the Nyquist
wave number for the chosen grid ($L$ is the size of the simulation
box). Finally, the power spectrum is computed by averaging modes with
wave numbers within narrow intervals $k-\bigtriangleup k/2< |\vec{k}|<
k+\bigtriangleup k/2$

\begin{equation}
P(k)=\frac{1}{\rm N_{modes}} \sum_{\vec{k}\in k} \delta(\vec{k}) \delta^*(\vec{k})~,
\end{equation}
where ${\rm N_{modes}}$ is the number of modes in the interval over
which $P(k)$ is computed. In practice, we compute the power 
spectrum in bins of size $2\pi/L$, using a grid with $N=512$. 

The previous procedure is used to compute the power spectrum of
the CDM particles, the neutrinos and the dark matter haloes. 
When the computational box contains more than one mass component, 
e.g. CDM and neutrinos, the power spectrum of the total mass is obtained 
from the following density field:
\begin{equation}
\delta_{\rm m}(\vec{r})=\frac{\Omega_{\rm cdm}}{\Omega_{\rm cdm}+ \Omega_\nu}\delta_{\rm cdm}(\vec{r})+\frac{\Omega_\nu}{\Omega_{\rm cdm}+\Omega_\nu}\delta_\nu(\vec{r})~,
\end{equation}
where $\delta_{\rm cdm}(\vec{r})$ and $\delta_\nu(\vec{r})$ are the
values of the density contrast in the mesh points for the CDM and the
neutrinos respectively. For the cross-spectrum of two different fluids
we compute the values of density contrast in Fourier space for each
fluid, $\delta_1(\vec{k})$ and $\delta_2(\vec{k})$, correcting for the
CIC procedure as outlined above, and we compute their cross-spectrum
by calculating:
\begin{equation}
P_{12}(k)=\frac{1}{\rm N_{modes}} \sum_{\vec{k}\in k} {\rm Re}\big[\delta^*_1(\vec{k}) \delta_2(\vec{k})\big]~.
\end{equation}

The measurements of the auto-power and cross-power spectra may require
a correction to account for the discreteness nature of the objects
used to compute them. If so, the measured halo power
spectrum will be shifted with respect to the correct one by the
inverse of the halo number density, $\bar{n}_{\rm haloes}^{-1}$, on
all scales (this is called white noise, since it is independent of
scale). We have corrected our power spectrum measurements to account
for this effect. The details of this correction for the different
cases can be found on the appendix \ref{Appendix_A}.

\begin{figure}
\begin{center}
\includegraphics[width=1\textwidth]{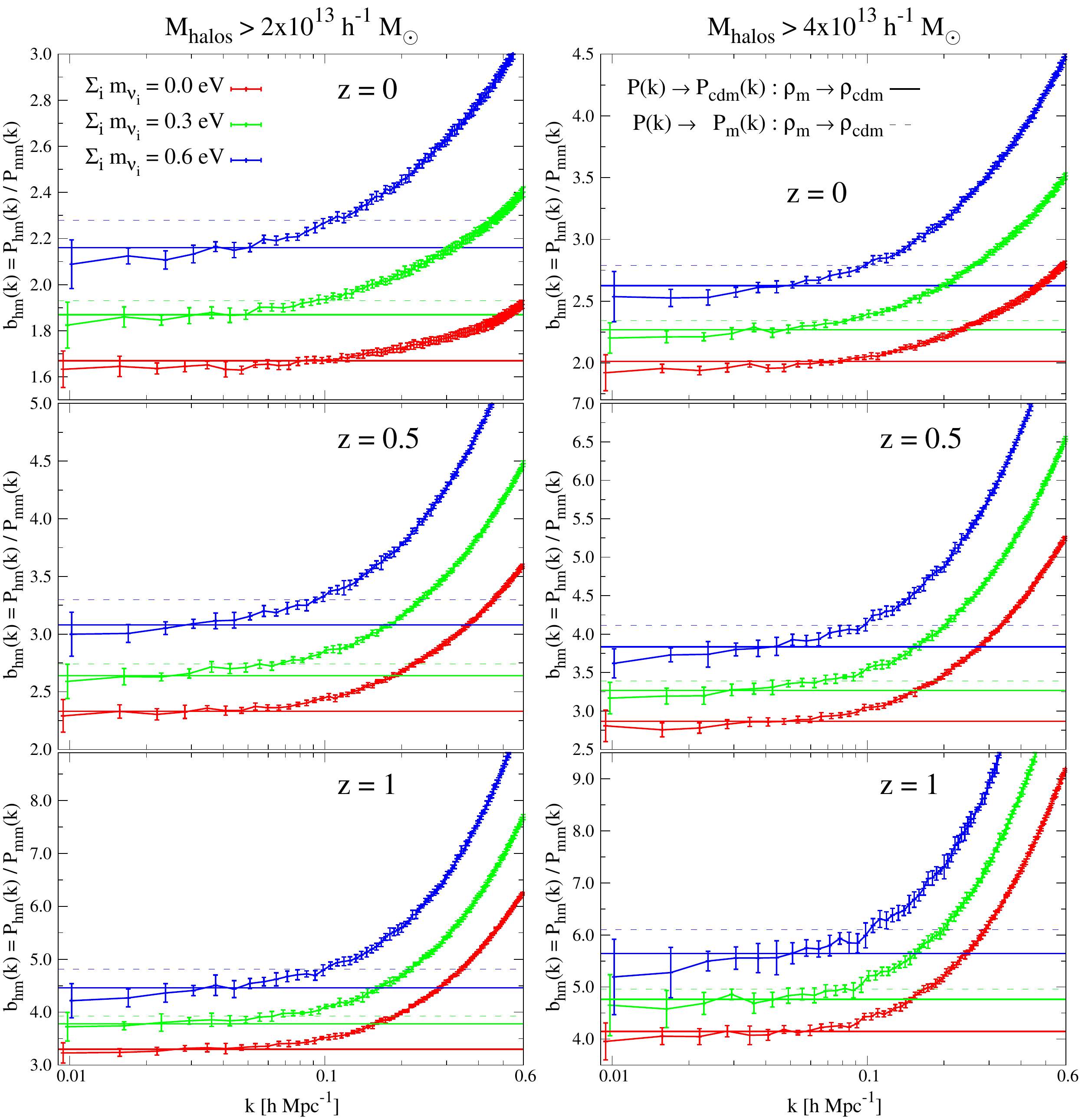}
\end{center}
\caption{Halo-matter bias, computed as the ratio of the
   halo-matter cross-power spectrum to the matter power spectrum, for
  massless and massive neutrino simulations. We show the bias for
  haloes with masses, $M_{200}$, larger than $2\times10^{13}~h^{-1}{\rm M}_\odot$
  (left column) and for haloes with masses larger than
  $4\times10^{13}~h^{-1}{\rm M}_\odot$ (right column) at redshifts
  $z=0$, $z=0.5$ and $z=1$ (as indicated on each panel). The
  results are shown for
  three different cosmologies with the same value of $\Omega_{\rm
    m}=0.2708$ and $A_s$ but different neutrino masses: \Mnu = 0.0 eV
  (red), \Mnu = 0.3 eV (green) and \Mnu = 0.6 eV (blue). The points and the error bars
  represent the mean value and the standard deviation from the set 
  of 8 independent realizations comprising each cosmological model. The
  horizontal lines show the value of the bias on large scales as
  predicted by the Tinker \cite{Tinker_2010} fitting formula. We have
  used two different prescriptions when using that formula: the
  \textit{matter prescription} (thin dashed lines), consisting in
  using the linear power spectrum for the whole matter and the
  \textit{cold dark matter prescription} (thick solid lines), in which
  we use the linear matter power spectrum of the CDM component. In
  both prescriptions we take $\rho_{\rm cdm}$ as the value of
  $\rho_{\rm m}$. Note that the results for the \Mnu = 0.3 eV and
  \Mnu = 0.6 eV cosmologies have been slightly displaced in the
  $x-$axis for clarity.}
\label{bias_hm}
\end{figure}

We use the simulations H0, H3 and H6 to study the impact of 
neutrino masses on the spatial distribution of dark
matter haloes. Those
simulations have the same cosmological parameter values, with the
exception of $\Omega_\nu$, $\Omega_{\rm cdm}$ and $\sigma_8$. 
The value of $\Omega_{\rm m}=\Omega_{\rm cdm}+\Omega_{\rm b}+\Omega_{\rm
  b}+\Omega_\nu$ is held fixed at 0.2708 for the three simulations,
 therefore, by comparing the
results among the different models, we are comparing universes in
which the same dark matter content is made up by different fractions
of massive neutrinos. Moreover, the amplitude of the power spectrum is
fixed on large scales, implicitly assuming that those scales are very
well constrained by the CMB data. This constraint implies that
the value of $\sigma_8$ depends on the neutrino masses, in such a way
that the larger the neutrino masses the lower the value of
$\sigma_8$. Each simulation consists of 8 independent realizations of
the same density field, obtained by using a different seed when
computing the initial conditions. For each realization, we select the
dark matter haloes with masses, $M_{200}$, above
$2\times10^{13}~h^{-1}{\rm M}_\odot$ and $4\times10^{13}~h^{-1}{\rm
  M}_\odot$ and compute the haloes power spectrum, the
  haloes-dark matter cross-power spectrum,
the matter-matter auto-power spectrum and the bias: $b_{\rm
  hm}(k)=P_{\rm hm}(k)/P_{\rm mm}(k)$ and $b_{\rm
  hh}^2(k)=P_{\rm hh}(k)/P_{\rm mm}(k)$. We have selected the
dark matter haloes with those masses to, on one hand, maximize
our statistics while, on the other hand, being able to investigate
whether our results are sensitive to the halo masses.
  
In Fig. \ref{bias_hm} we show, for each cosmological model, 
the value of 
$b_{\rm hm}(k)$, averaged 
over the 8 realizations and 
its r.m.s. scatter (errorbars). The blue points represent the results for the \Mnu =
0.6 eV cosmology, whereas the green and red points are for the \Mnu =
0.3 and \Mnu = 0.0 eV cosmologies, respectively. The results are shown
at $z=0$, $z=0.5$ and $z=1$ for the two mass
thresholds: $2\times10^{13}~h^{-1}{\rm M}_\odot$ (left column) and
$4\times10^{13}~h^{-1}{\rm M}_\odot$ (right column). We do not show 
the results at $z = 2$ since the number of haloes above threshold is small 
and results are very noisy.

We find that at any given wave number, the bias increases with the
neutrino masses. The different value of the parameter $\sigma_8$ 
in the three cosmological models considered is the main reason of this
effect. Objects with masses larger than
$2\times10^{13}~h^{-1}{\rm M}_\odot$ are rarer in a cosmology with
$\sigma_8=0.675$ (as in the cosmology with \Mnu = 0.6 eV) than in a
cosmology with $\sigma_8=0.832$ (as in the massless neutrino
cosmology), and thus, the bias is expected to be higher in the former model
than in the latter \cite{Marulli_2011}. On large scales, we find
that the bias flattens out for all the cosmological models. However,
we find that even on large scales the bias displays some weak
scale-dependence for the simulations with massive neutrinos. We will
discuss this point in detail below.

We check whether our results on large scales can reproduced by the 
Tinker fitting formula \cite{Tinker_2010}
\begin{equation}
b_{\rm Tinker}(\nu)=1-A\frac{\nu^a}{\nu^a+\delta_c^a}+B\nu^b+C\nu^c~,
\end{equation}
with $A=1.0+0.24ye^{-(4/y)^4}$, $a=0.44y-0.88$, $B=0.183$, $b=1.5$,
$C=0.019+0.107y+0.19e^{-(4/y)^4}$ and $c=2.4$, being $y={\rm log}_{10}
\bigtriangleup$. $\bigtriangleup$ is the halo mean overdensity within
its virial radius in units of the mean matter density of the universe,
which is set to 200 for this paper. In the Tinker fitting formula, the 
dependence of the bias on the halo mass is parametrized in terms of the
\textit{peak height} of the linear density field:
$\nu=\delta_c/\sigma(M)$, where $\delta_c=1.686$\footnote{We have
  neglected the dependence of this parameter with cosmology and with
  the masses of the neutrinos since that is very weak.} is the value of the linearly
extrapolated overdensity at the time of collapse and $\sigma(M)$ is
defined as

\begin{equation}
\sigma^2(M)=\frac{1}{2\pi^2}\int_0^\infty P_{\rm lin}(k,z)\hat{W}(k,R)k^2dk~,
\end{equation}
where $P_{\rm lin}(k,z)$ is the linear matter power spectrum at
redshift $z$ and $\hat{W}(k,R)$ is the Fourier transform of the top-hat
window function of radius $R$. The relationship between the halo mass
and the variable $R$ is given by $M=4\pi\rho_{\rm m}R^3/3$, with
$\rho_{\rm m}$ being the mean matter density of the universe.

When using the Tinker fitting formula we need to specify both the
linear power spectrum $P_{\rm lin}(k)$ and $\rho_{\rm m}$; the latter
relates the radius in the top-hat window function
to the halo mass, whereas the former enters in
the definition of $\sigma(M)$. For 
cosmologies with massless neutrinos the linear power spectrum of the 
cold dark matter and that of the total matter coincide. However, when 
more components contribute to $\Omega_{\rm m}$ it is not clear
which power spectrum is more relevant to describe halo clustering and their abundance.
Recent studies related to the impact of massive neutrinos on the halo mass
function \cite{Brandbyge_haloes, Marulli_2011, Villaescusa-Navarro_2013a} 
have suggested that a reasonable
agreement between theory and simulations is achieved if the total
matter linear power spectrum is used, together with $\rho_{\rm
m}=\rho_{\rm cdm}$ (used to establish the M-R relationship). 
In this paper we use that prescription for massive neutrinos cosmologies,
that we will refer to as the \textit{matter prescription}, when
computing the value of the bias and the halo mass function. Moreover, we
also test the \textit{cold dark matter prescription}, which uses the
linear power spectrum of the CDM component together with $\rho_{\rm
  m}=\rho_{\rm cdm}$, as suggested by \cite{Ichiki-Takada} and tested
in our companion papers \cite{Castorina, Costanzi2} agains N-body simulations 
for studies of the halo mass functions in universes with massive neutrinos.

We compute the linear power spectrum for the CDM component
through CAMB as:
\begin{equation}
P_{\rm lin}^{\rm cdm}(k)=\left(\frac{T_{\rm cdm}}{T_{\rm m}}\right)^2P_{\rm lin}^{\rm m}(k)~,
\end{equation}
where the $T_{\rm cdm}(k)$\footnote{In our case we incorporate the
baryons effects into the CDM particles when generating the initial conditions by setting:
$T_{\rm cdm}(k)=[\Omega_{\rm
      cdm}\,T_{\rm cdm}(k)+\Omega_{\rm b}\,T_{\rm b}(k)]/(\Omega_{\rm cdm}+\Omega_{\rm b})$,
  where $T_{\rm cdm}(k)$ and $T_{\rm b}(k)$ are directly obtained via
  CAMB.} and $T_{\rm m}(k)$ are the transfer functions for the CDM
and the total matter respectively. To compare our results with the fitting
formula of Tinker, we need to calculate the effective bias for the
haloes we selected in our simulations, i.e. for haloes with masses above a 
certain mass
\begin{equation}
\bar{b}(z)=\frac{\int_{M_{\rm min}}^{M_{\rm max}}b_{\rm Tinker}(M,z)n(M,z)dM}{\int_{M_{\rm min}}^{M_{\rm max}}n(M,z)dM}~,
\end{equation}
where $M_{\rm min}$ and $M_{\rm max}$ are the minimum and maximum
masses of the dark matter haloes used in the analysis. The halo mass
function, $n(M,z)dM$, specifies the comoving number density of dark matter haloes
with masses within the interval $[M,M+dM]$. For the halo mass function we
use the fitting formula of Tinker et al. 2008 \cite{Tinker_2008}
with the same prescriptions as above for cosmologies with massive neutrinos,
 i.e. $\sigma(M)$ is computed
using the total matter linear power spectrum when using the \textit{matter
  prescription}, whereas the linear CDM power spectrum is used for the
\textit{cold dark matter prescription}. In both prescriptions we use
$\rho_{\rm cdm}$ as the value of $\rho_{\rm m}$. In \cite{Costanzi2} we show that
the halo mass function of SO haloes in cosmologies with massive neutrinos is well
reproduced by the Tinker fitting formula along with the 
\textit{cold dark matter prescription}. We set 
$M_{\rm max}=3\times10^{15}~h^{-1}{\rm M}_\odot$.
We have explicitly checked that our results do not
significantly change if we choose a different (but reasonable) value
of $M_{\rm max}$. 

In Fig. \ref{bias_hm} the horizontal, thin dashed line shows the prediction
of Tinker's bias obtained using the \textit{matter prescription}, as opposed to the 
thick, solid horizontal lines that refer to the \textit{cold dark matter prescription}.
We find that, at all
redshifts and for the two different mass thresholds, the bias on large
scales is better reproduced if we use the Tinker fitting formula
 with the \textit{cold dark matter prescription} for massive
neutrinos. This is more clearly seen in the results for the \Mnu = 0.6
eV cosmology, where the value of the bias on large scales computed
using the \textit{matter prescription} is shifted to higher values
than the ones obtained from the N-body simulations. This point is 
investigated more extensively in our companion paper \cite{Castorina}.

\begin{figure}
\begin{center}
\includegraphics[width=1\textwidth]{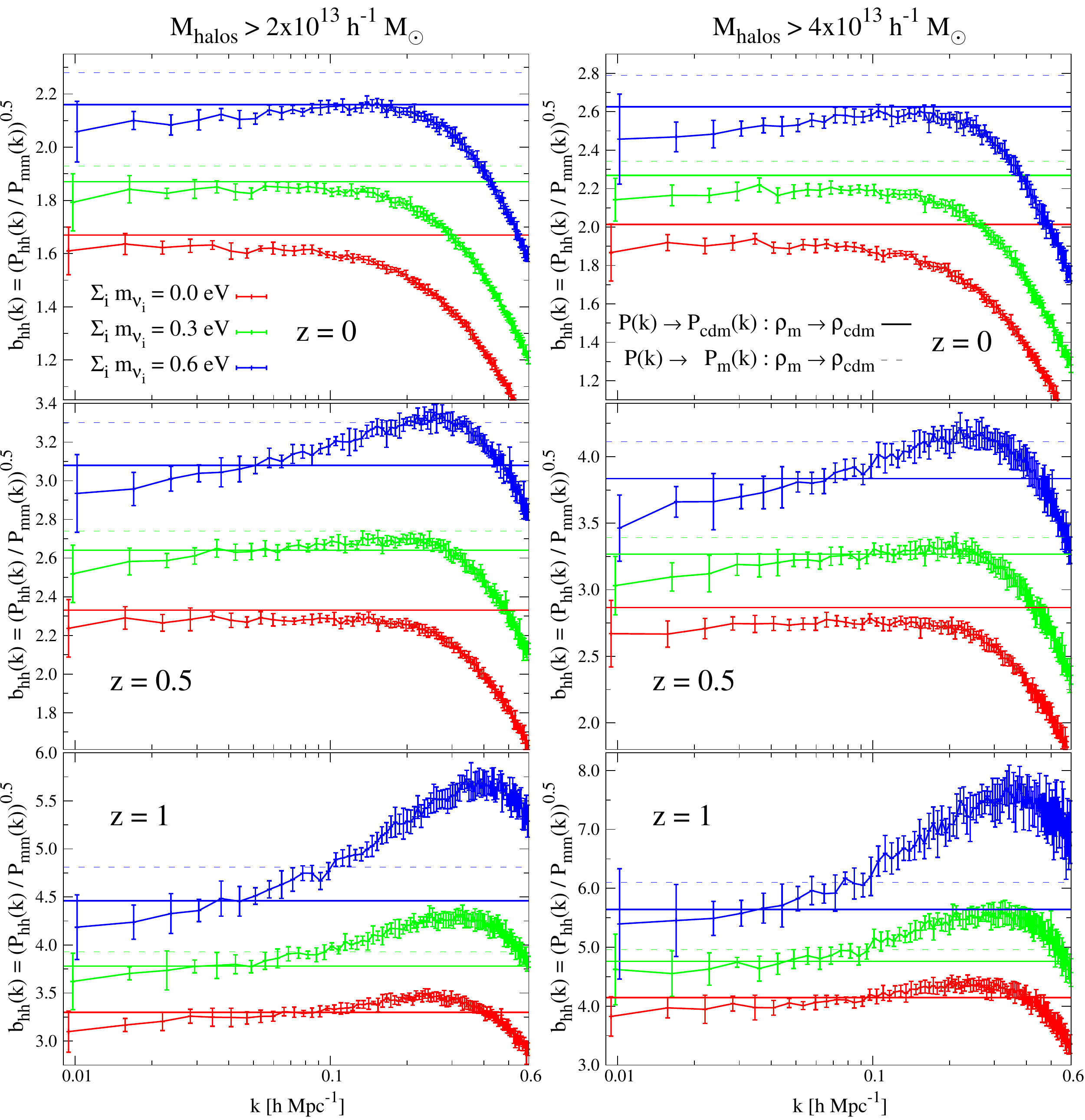}
\end{center}
\caption{Same as Fig. \ref{bias_hm} but with the bias computed as $b_{\rm hh}^2(k)=P_{\rm hh}(k)/ P_{\rm mm}(k)$.}
\label{bias_hh}
\end{figure}

In Fig. \ref{bias_hh} we show the results when the bias is computed
using the definition $b_{\rm hh}^2(k)=P_{\rm hh}(k) / P_{\rm
  mm}(k)$. In this case as well we find that the bias of the
dark matter haloes increases with the neutrino masses, as a
consequence of the different values of the parameter $\sigma_8$ for
the models studied here. In contrast to the results of the bias
computed using the halo-matter cross-power spectrum, $b_{\rm hm}(k)$,
we find that the value of bias, $b_{\rm hh}(k)$, drops on small
scales.  This is due to the halo-exclusion effect: the probability of
finding two dark matter haloes on very small scales shrinks quickly,
as a consequence of the finite size of those. We also find that the
Tinker fitting formula reproduces reasonably well our results, in
terms of the bias on large scales when the \textit{cold dark matter prescription}
for cosmologies with massive neutrinos is used. The disagreement between 
our results and the Tinker fitting formula is larger is the latter is used along the 
\textit{matter prescription}.  However,
we note that the agreement between our results on large scales and the
Tinker fitting formula plus the \textit{cold dark matter prescription}
is poorer than when the bias is computed as
$b_{\rm hm}(k)=P_{\rm hm}(k)/ P_{\rm mm}(k)$. We also find this effect when
the bias is measured using the correlation function (see the following
subsection). The scale-dependent bias on large scales is clearly seen
in the models with massive neutrinos, being more prominent at high
redshift and for the model with \Mnu = 0.6 eV neutrinos.

\begin{figure}
\begin{center}
\includegraphics[width=1\textwidth]{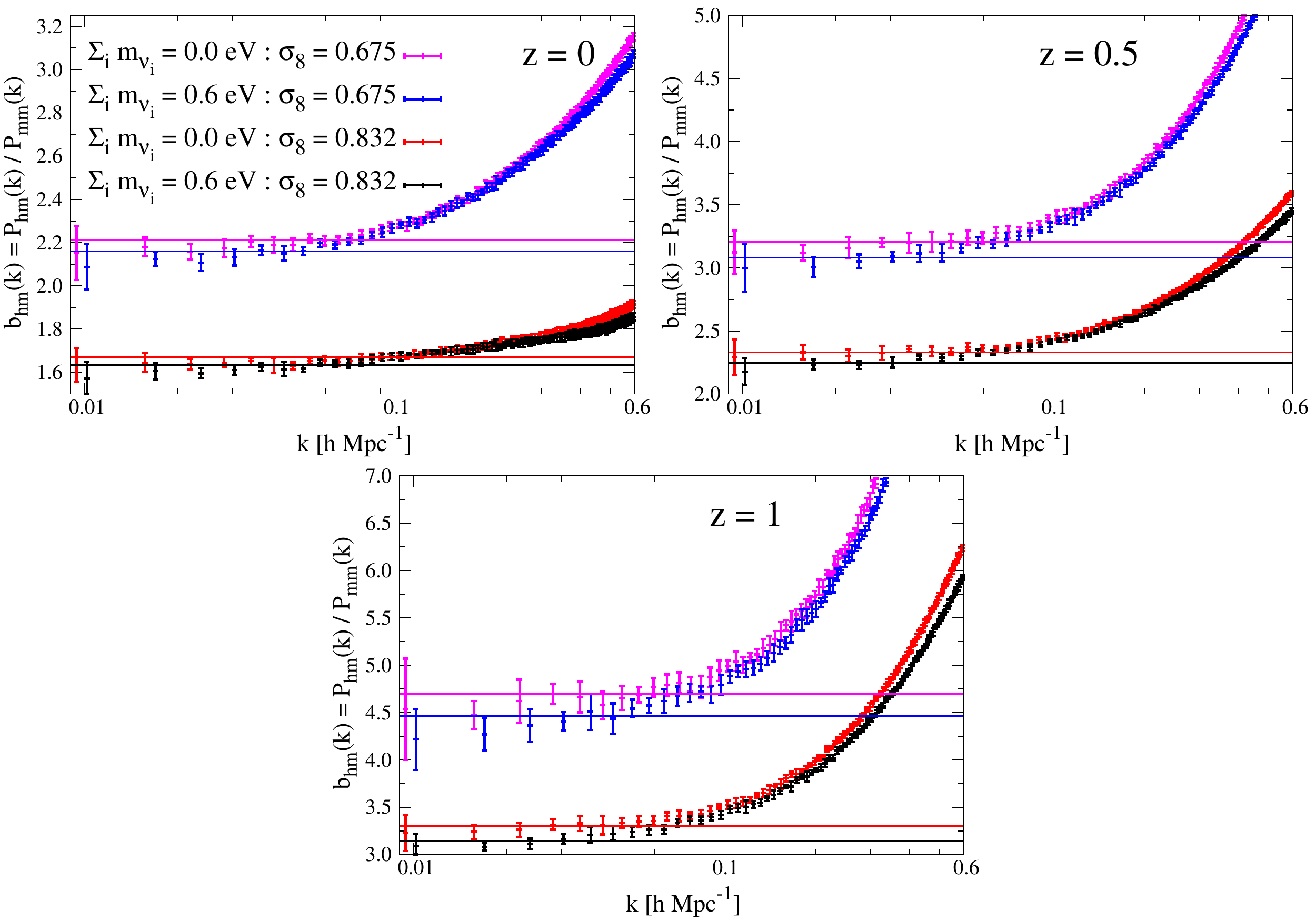}
\end{center}
\caption{$\Omega_\nu-\sigma_8$ degeneracy in the halo-matter bias. We
  plot the bias, $b_{hm}$, between the spatial distribution of haloes with
  masses, $M_{200}$, above $2\times10^{13}~h^{-1}{\rm M}_\odot$, and
  this of the underlying dark matter, comparing different cosmologies
  with the same value of $\sigma_8$ but different \Mnu. The bias is
  shown at three different redshifts: $z=0$ (top-left), $z=0.5$
  (top-right) and $z=1$ (bottom). Models
  with a value of $\sigma_8=0.675$ are represented by the magenta and
  the blue points, which have $\Sigma_i m_{\nu_i}=0$ eV and $\Sigma_i
  m_{\nu_i}=0.6$ eV respectively. The red and black lines correspond
  to models with \Mnu= 0 eV and \Mnu= 0.6 eV respectively, with the
  same value of $\sigma_8=0.832$. The horizontal lines show the
  prediction of the bias on large scales using the Tinker formula plus
  the \textit{cold dark matter prescription} for neutrinos. The error
  bars are the dispersion around the mean bias value from the set of 8
  independent realizations comprising each simulation.}
\label{bias_Pk_Mnu-s8}
\end{figure}

We now focus on the $\Omega_\nu-\sigma_8$ degeneracy.  Our purpose
here is to investigate how the spatial distribution of dark matter haloes is
affected by \Mnu for cosmological models that share the same value of
the cosmological parameter $\sigma_8$. To answer this question, we
compare the bias computed from the simulations H0 and H6 with
the bias measured in the simulations H6s8 and H0s8. The former
represents a cosmology with the same value of $\sigma_8$ as the
simulation H0, but it also contains neutrinos with masses equal to
\Mnu= 0.6 eV, whereas the latter corresponds to a cosmology with
massless neutrinos but with the same value of $\sigma_8$ as the 
simulation H6. In Fig. \ref{bias_Pk_Mnu-s8} we show the results, in terms of 
$b_{\rm hm}(k)$, together with the
Tinker prediction using the \textit{cold dark matter prescription} for massive
neutrinos. The error bars represent the standard deviation around the mean
value from the set of 8 realizations available for each cosmological
model. We find that the bias for simulations that
share the same value of $\sigma_8$ are very similar, though not
exactly the same. On large scales, our results are well reproduced by
the fitting formula of Tinker along with the \textit{cold dark matter
prescription} for models with massive neutrinos. For models with the same value of
$\sigma_8$ the Tinker fitting formula predicts a difference on the bias of
$\sim3\%$ at $z=0$ while this number increases up to $\sim8\%$ at
$z=2$. It is not surprising that the differences between models that
share the same value of $\sigma_8$ increase with redshift, since their
growth factors are different. Given these
results we conclude that the $\Omega_\nu-\sigma_8$ degeneracy (in
terms on the bias on large scales) is not perfect. In terms of the Tinker
fitting formula, the value of large-scale bias will be degenerate
in \Mnu for models that share the same linear CDM power spectrum. We
further discuss this point in our companion paper \cite{Castorina}.

\begin{figure}
\begin{center}
\includegraphics[width=1\textwidth]{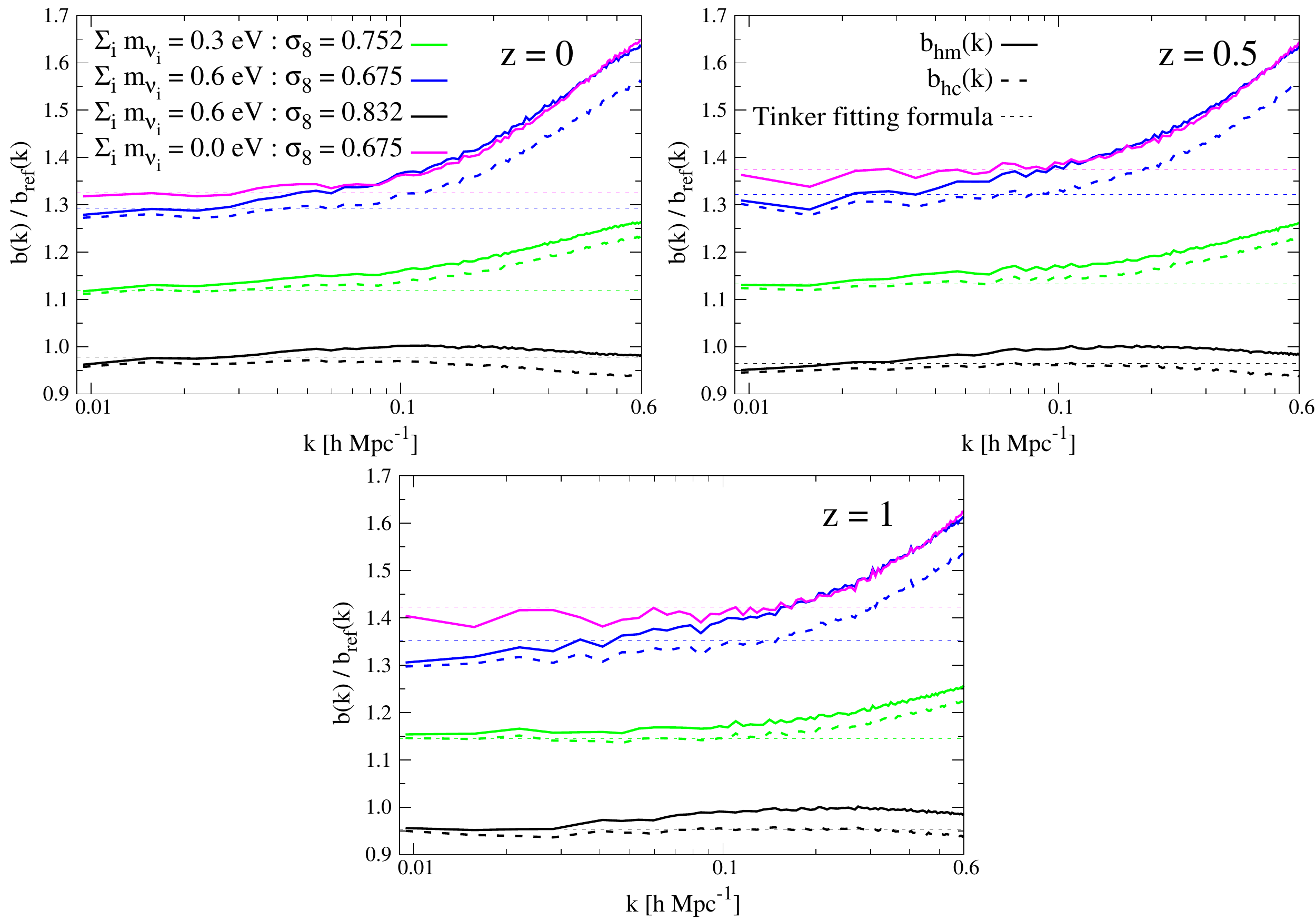}
\end{center}
\caption{Bias ratio for different cosmological models. We show the
  bias of dark matter haloes with masses above
  $2\times10^{13}~h^{-1}{\rm M}_\odot$ for the simulations H6 (blue),
  H3 (green), H6s8 (black) and H0s8 (magenta) normalized by the bias
  of the simulation H0 (reference model). The results are shown at
  four different redshifts, whose value is indicated on each
  panel. The solid lines show the results when the bias is computed
  using the total matter distribution, i.e. CDM plus massive
  neutrinos, whereas the dashed lines represent the results when the
  bias is calculated by using just the CDM distribution. In our
  companion paper \cite{Castorina}, we show that only by computing the
  bias with respect to the cold dark matter distribution, the bias on
  large scale becomes scale-independent and universal.}
\label{ratio_Pk}
\end{figure}

The large scale dependence of the bias on scales $k<0.1~h~{\rm Mpc}^{-1}$ 
in presence of massive neutrino deserves some additional investigation.
No such dependence is found when considering massless 
neutrinos (H0 and H0s8 simulations), as expected. We
emphasize this by showing with solid lines in Fig.  \ref{ratio_Pk} the
bias, $b_{\rm hm}(k)$, for the different cosmological models
normalized to the bias of the model with massless neutrinos and
$\sigma_8=0.832$ (simulation H0, reference model). 
The ratios are shown at redshifts
$z=0$, $z=0.5$ and $z=1$ for the simulations H6 (blue), H3 (green),
H6s8 (black) and H0s8 (magenta). The horizontal dotted lines represent
the Tinker prediction along with the \textit{cold dark matter
  prescription} for massive neutrinos. Focusing on the results of the
simulations H3 and H6 (sharing the same value of $A_s$ as
the reference model), we find a clear scale-dependent bias trend on
large scales. If we compare the results of the simulation H0s8
(magenta) with the one of the simulation H6 (blue), both sharing the same
value of the parameter $\sigma_8$, we find that for the former, the
bias is constant for $k\lesssim0.1~h~{\rm Mpc}^{-1}$,
whereas we find a clear scale-dependent bias in the simulation with
\Mnu = 0.6 eV neutrinos (H6). In our companion paper \cite{Castorina},
we show that this effect is due to the presence of massive neutrinos,
and that only by calculating the bias as the ratio $P_{\rm
  hc}(k)/P_{\rm cc}(k)$ where $P_{\rm hc}(k)$ and $P_{\rm cc}(k)$ are
the haloes-cold dark matter and cold dark matter power spectrum
respectively, the bias becomes scale-independent, and universal, on
large scales. We have computed the bias using $P_{\rm hc}(k)/P_{\rm
  cc}(k)$ for the models with massive neutrinos, and we show the
results of the ratio with respect to the reference model in
Fig. \ref{ratio_Pk} with dashed lines. On large scales
($k\lesssim0.1~h^{-1}$Mpc) we find that the scale-dependence of the
bias is highly suppressed if quantities are computed with respect to
the CDM distribution instead of over the total matter distribution.

\subsection{Halo bias: correlation function}
\label{subsec:correlation_function}

In this subsection we further investigate the clustering properties of dark
matter haloes in cosmologies with massive neutrinos using the correlation
function statistics, i.e. the
Fourier transform of the power spectrum. While the information content 
is in principle the same as with the power spectra, the use of a different 
estimate allows to emphasize clustering properties on different scales.
The purpose of this section is, thus, to corroborate the
results obtained in the previous section.

As we have done in the above subsection with the power spectrum technique, 
we study the impact of massive neutrinos on the bias between the
spatial distribution of dark matter haloes and that of the underlying
matter using two different estimators:
$b_{\rm hh}^2(r)=\xi_{\rm hh}(r)/\xi_{\rm mm}(r)$ and 
$b_{\rm hm}(r)=\xi_{\rm hm}(r)/\xi_{\rm mm}(r)$, where
$\xi_{\rm hh}(r)$, $\xi_{\rm hm}(r)$ and $\xi_{\rm mm}(r)$ 
are the 2-pt autocorrelation functions of the dark matter haloes, the 
haloes-matter cross-correlation function and the matter 2-pt autocorrelation
function, respectively. 

Our halo catalogue consists of all the dark matter haloes with masses, $M_{200}$,
above $2\times10^{13}~h^{-1}{\rm M}_\odot$, as given by the SUBFIND groups.
The matter 2-pt autocorrelation function and the haloes-matter cross-correlation function is not
measured using all the dark matter particles (i.e. CDM particles plus neutrinos) since the 
computational cost of such calculation is prohibitively large. We however decide 
to randomly select 5 million particles of both types.
We have checked that our results are robust to this choice in the sense 
that increasing the number of randomly selected 
particles by a factor  of 2 does not significantly change them.

The 2-pt autocorrelation functions are computed using the
Landy-Szalay estimator \cite{Landy-Szalay_93}:
\begin{equation}
\xi(r)=\frac{{\rm DD}(r)-2{\rm DR}(r)+{\rm RR}(r)}{{\rm RR}(r)}~.
\end{equation}
Here ${\rm DD}(r)$ and ${\rm RR}(r)$ are the normalized
number of pairs with distances in the interval $[r,r+dr]$, in the
initial catalogue (data-data) and in the so-called random catalogue
(random-random) respectively, whereas ${\rm DR}(r)$ represents the 
normalized number of cross-pairs between both catalogues
(data-random). We use the above estimator to compute 
the 2-pt autocorrelation functions of the dark matter haloes, the CDM
particles and the neutrinos.

The cross-correlation function between 
objects of type A and objects of type B is computed using the following 
estimator \cite{Szapudi}:
\begin{equation}
\xi_{\rm AB}(r)=\frac{{\rm D}_{\rm A}{\rm D}_{\rm B}(r)-{\rm
    D}_{\rm A}{\rm R}(r)-{\rm D}_{\rm B} {\rm R}(r)+{\rm RR}(r)}{{\rm
    RR}(r)}~,
\label{cross_CF_estimator}
\end{equation}
where ${\rm D}_{\rm A}{\rm D}_{\rm B}(r)$, ${\rm D}_{\rm A}{\rm R}(r)$,
${\rm D}_{\rm B} {\rm R}(r)$ and ${\rm RR}(r)$ are the normalized number of
type A objects-type B objects, type A objects-random, type B objects-random
and random-random pairs with distances in the interval $[r,r+dr]$. We use
this estimator to compute the CDM-neutrino, CDM-haloes and neutrino-haloes 
cross correlation functions.

In both estimators the random catalogue is constructed by filling the simulation box
with 10 million particles randomly distributed within the interior
volume. We have explicitly checked that our results do not change if we increase
the number of random points when constructing the random
catalogue. When computing the distances among pairs, we have taken
into account the periodic boundary conditions of the simulation
box.

\begin{figure}
\begin{center}
\includegraphics[width=0.49\textwidth]{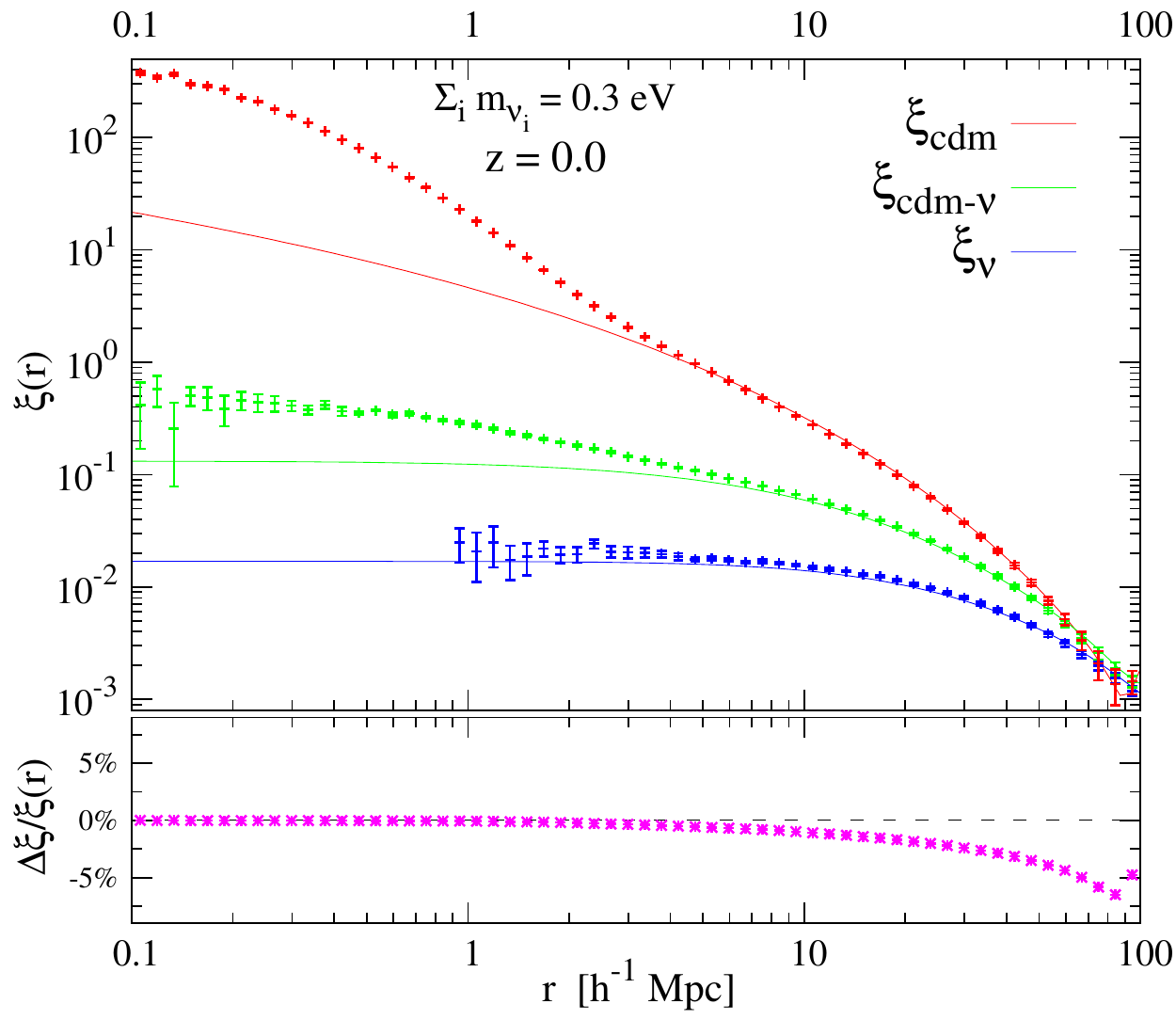}
\includegraphics[width=0.49\textwidth]{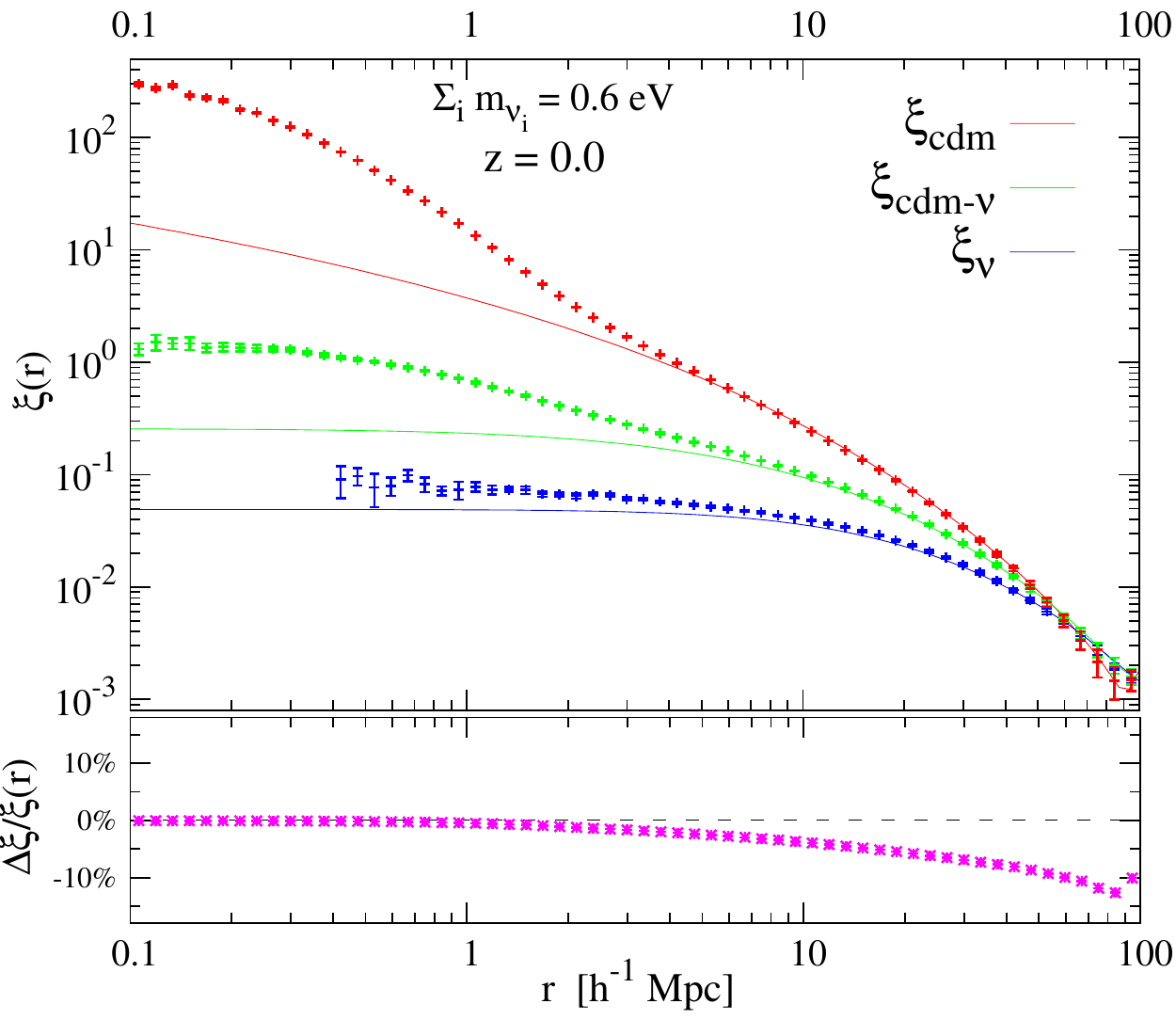}
\end{center}
\caption{The panels show the 2-pt autcorrelation function of the CDM (red),
  the neutrinos (blue) and their cross-correlation (green) for the
  cosmologies with \Mnu= 0.3 eV (left panel) and \Mnu= 0.6 eV (right
  panel) at $z=0$. The thin solid lines correspond to the
  linear theory prediction obtained from CAMB (see text for details). 
  The bottom panels show the relative difference between computing the 
  correlation function for the entire matter with and without the last two term in the
  right hand side of Eq. \ref{cross_correlation}, i.e. the
  contribution of massive neutrinos to the total matter correlation
  function.}
\label{TPCCF}
\end{figure}

If the N-body simulation contains more than one particle type, i.e. CDM and neutrinos, we  compute the correlation function of the total matter distribution by calculating 
\begin{equation}
\xi_{\rm m}(r)=\left(\frac{\Omega_{\rm cdm}}{\Omega_{\rm m}}\right)^2\xi_{\rm cdm}(r)+
\left(\frac{\Omega_\nu}{\Omega_{\rm m}}\right)^2\xi_{\nu}(r)+\left(\frac{2~\Omega_{\rm cdm}~\Omega_\nu}{\Omega_{\rm m}^2}\right)\xi_{\rm cdm-\nu}(r)~,
\label{cross_correlation}
\end{equation}
being $\Omega_{\rm m}=\Omega_{\rm cdm}+\Omega_{\rm b}+\Omega_\nu$. $\xi_{\rm cdm}(r)$ and $\xi_\nu(r)$ are the 
cold dark matter and neutrino 2-pt autocorrelation functions and $\xi_{\rm cdm-\nu}(r)$ is the 
cross-correlation among them. The haloes-matter cross-correlation function in simulations with
massive neutrinos is computed as
\begin{equation}
\xi_{\rm hm}(r)=\left(\frac{\Omega_{\rm cdm}}{\Omega_{\rm m}}\right)\xi_{\rm hc}(r)+
\left(\frac{\Omega_\nu}{\Omega_{\rm m}}\right)\xi_{\rm h\nu}(r)~,
\label{cross_correlation_hm}
\end{equation}
with $\xi_{\rm hc}(r)$ and $\xi_{\rm h\nu}(r)$ being the haloes-CDM and the haloes-neutrinos
cross-correlation functions computed using the estimator \ref{cross_CF_estimator}.

We now study the impact of massive neutrinos on the bias between the
spatial distribution of dark matter haloes and that of the underlying
matter using the models H0, H3 and H6 sharing
the value of $\Omega_{\rm m}$ and $A_s$,
i.e. models in which
neutrinos made up a different fraction of the same dark matter content
and models in which the amplitude of the power spectrum is fixed on
large scales (error bars from CMB measurements are small on those
scales). For each 
realization we measure the 2-pt autocorrelation functions of the dark matter
haloes, the CDM particles, the neutrinos and 
the total matter, together with
the CDM-neutrinos, CDM-haloes, neutrino-haloes and matter-haloes 
cross-correlation functions using the above method and estimators.

\begin{figure}
\begin{center}
\includegraphics[width=1.0\textwidth]{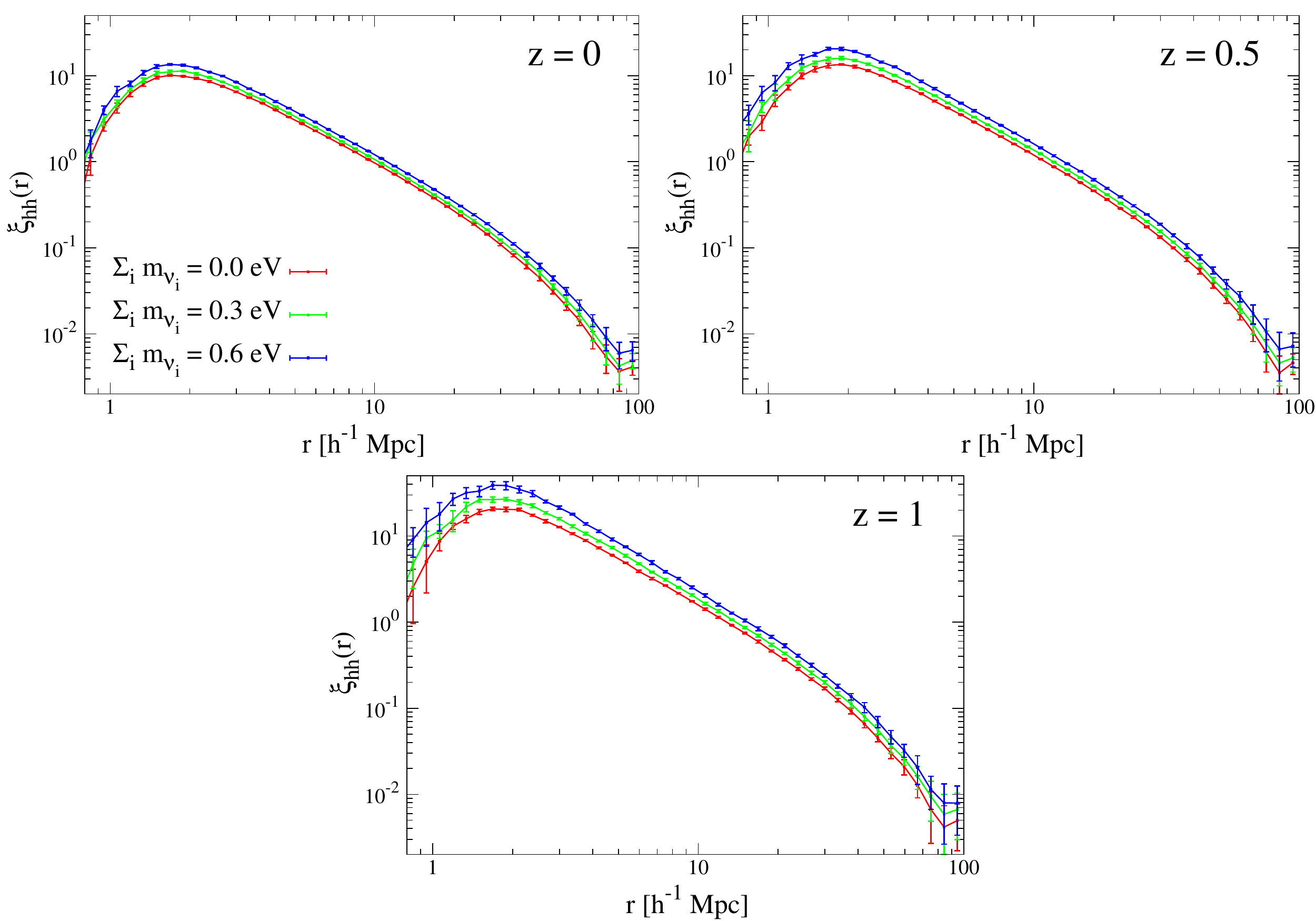}
\end{center}
\caption{Dark matter haloes autocorrelation function. We show the correlation function of dark
matter haloes with masses, $M_{200}$, larger than $2\times10^{13}~h^{-1}{\rm M}_\odot$
for three cosmological models: \Mnu = 0.0 eV (red), \Mnu = 0.3 eV (green) and \Mnu = 0.6 eV (blue).
Results are displayed at $z=0$ (top-left), $z=0.5$ (top-right) and $z=1$ (bottom). The points and
error bars represent the mean and the dispersion around the mean obtained from the set 
of 8 independent realizations comprising each simulation.}
\label{halo_CF}
\end{figure}

In Fig. \ref{TPCCF} we show the 2-pt autocorrelation function of the CDM
and neutrino components together with their cross-correlation function in real
space, at $z=0$, for the cosmologies with \Mnu = 0.3 eV (left panel)
and \Mnu = 0.6 eV (right panel).
 The points represent the mean over the 8 independent
realizations while the error bars are the standard deviation around the
mean. The thin solid lines represent the linear (cross-)correlation function
of each object, obtained from the linear power spectrum [as given by
CAMB. For each object, its linear power spectrum has been computed
using the CAMB transfer functions, and their correlation functions are
calculated as follows:
\begin{equation}
\xi(r)=\frac{1}{2\pi^2}\int_0^\infty P_{\rm lin}(k)\frac{\sin(kr)}{kr}k^2dk~.
\end{equation}

\begin{figure}
\begin{center}
\includegraphics[width=1.0\textwidth]{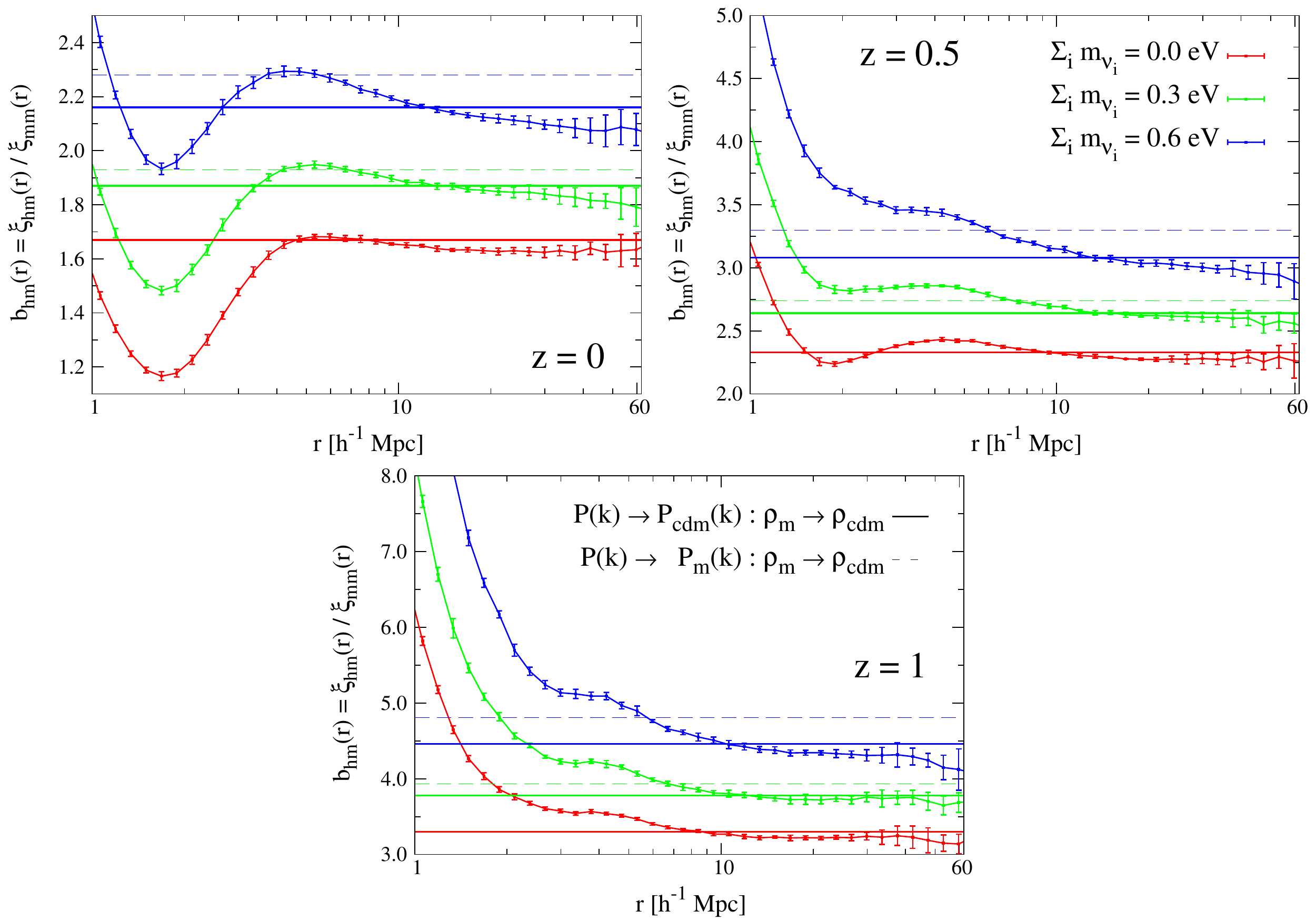}
\end{center}
\caption{Bias between the spatial distribution of dark matter haloes
  and this of the underlying matter, computed as $b_{\rm
    hm}(r)=\xi_{\rm hm}(r)/\xi_{\rm mm}(r)$. We show the results at
  $z=0$ (left), $z=0.5$ (middle) and $z=1$ (blue) for cosmological
  models that share the value of $\Omega_{\rm m}$ and $A_s$, but have
  different values of $\Omega_\nu$. The results for the \Mnu = 0 eV
  are shown in red, whereas the results for the models with \Mnu = 0.3
  eV and \Mnu = 0.6 eV are displayed in green and blue
  respectively. On large scales the bias flattens out, although we
  find the bias to be scale-dependent for the models with massive
  neutrinos. The horizontal lines represent the value of the bias on large scales
  obtained by using the Tinker fitting formula along with the 
\textit{cold dark matter prescription} (solid lines) and the \textit{matter prescription}.}
\label{bias_hm_CF_non_linear}
\end{figure}

We find that on scales larger than $\sim 20~h^{-1}$Mpc the correlation
function of each tracer (CDM, neutrinos or CDM-neutrinos) is
well described by the linear one. As expected, on small scales the nonlinear evolution of the
CDM density field is much stronger correlated with itself than the
neutrinos are. Their cross-correlation function indicates that
neutrinos are certainly correlated with the CDM density field (see for
instance \cite{Villaescusa-Navarro_2013a}). We also explore how
important are the corrections induced by neutrinos on the total matter
correlation function. We compute the matter 2-pt autocorrelation
function using $\xi_{\rm m}(r)=\left(\Omega_{\rm cdm}/\Omega_{\rm
  m}\right)^2\xi_{\rm cdm}(r)$, i.e. neglecting the neutrino
contribution, and compared it with the estimate from
Eq. \ref{cross_correlation}. In the lower panels of Fig. \ref{TPCCF}
we plot the relative difference between these quantities. 
We find that on scales smaller than
$\sim 5$ $h^{-1}$Mpc the contribution from neutrinos is indeed
negligible. This happens because on those scales the correlation
function is dominated by the 1-halo term, i.e. by the clustering of
matter within CDM haloes, and it is well known (see \cite{Ma, Wong,
  Brandbyge_haloes,Villaescusa-Navarro_2013a}) that neutrinos are
significantly less clustered than CDM on these scales.
 On the other hand, the contribution of
neutrinos, and in particular of the CDM-neutrino cross-correlation
function, becomes more important on large scales. At
$r\sim80~h^{-1}$Mpc, the neutrino contribution to the total matter
correlation function is larger than $5\%$ and $10\%$ for neutrinos
with \Mnu = 0.3 eV and \Mnu = 0.6 eV respectively.

In Fig. \ref{halo_CF} we show the autocorrelation function of the dark matter haloes with
masses, $M_{200}$, above $2\times10^{13}~h^{-1}{\rm M}_\odot$ for the three cosmological
model here considered at redshifts $z=0$, $z=0.5$ and $z=1$. On small scales we find that
the halo correlation function quickly drops as a consequence of the halo exclusion. 
We find that the scale where this phenomenon takes place is the same for all the cosmological
models.

\begin{figure}
\begin{center}
\includegraphics[width=1.0\textwidth]{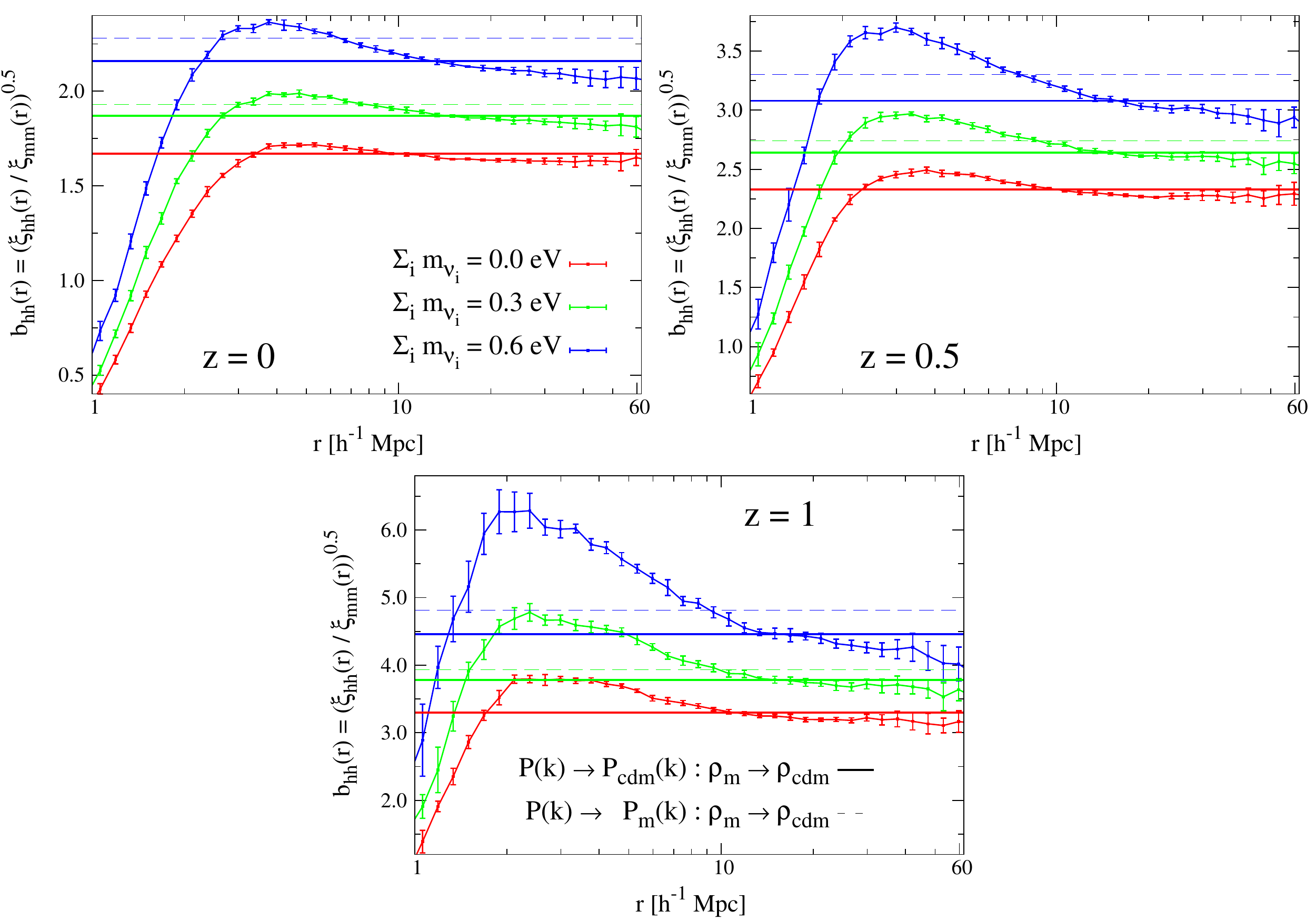}
\end{center}
\caption{Same as Fig. \ref{bias_hm_CF_non_linear} but with bias computed as 
$b_{\rm hh}^2(r)=\xi_{\rm hh}(r)/\xi_{\rm mm}(r)$.}
\label{bias_CF_non_linear}
\end{figure}

For each realization of the simulations H0, H3 and H6 we
compute the bias using the two definitions mentioned at the beginning of
this subsection: $b_{\rm hm}(r)=\xi_{\rm hm}(r)/\xi_{\rm mm}(r)$ and
$b_{\rm hh}(r)=\sqrt{\xi_{\rm hh}(r)/\xi_{\rm mm}(r)}$. In
Fig. \ref{bias_hm_CF_non_linear} we show the mean value of the bias, computed
as $b_{\rm hm}(r)$, and
its standard deviation for each cosmological model: \Mnu = 0.0 eV (red),
\Mnu = 0.3 eV (green) and \Mnu = 0.6 eV (blue). The results are shown
at redshifts $z=0$, $z=0.5$ and $z=1$. Again, we do
not show the results at $z=2$ because the number of haloes at this
redshift is very small, and results are very noisy. The transition
from the 2-halo term and the 1-halo term is clearly visible on small scales.
On large scales,
we find that the bias flattens out, approaching a constant
value. However, as in the above subsection, we find a weak
scale-dependence of bias on large scales for the cosmological models 
with massive neutrinos. With horizontal lines we show the value of the bias on large
scales obtained by using the Tinker fitting formula along the 
\textit{cold dark matter prescription} (solid lines) and the 
\textit{matter prescription} (dashed lines) for cosmologies with massive neutrinos. 
We find that the Tinker fitting formula plus the \textit{cold dark matter prescription}
reproduces reasonably well our results. As we found in the power spectrum subsection,
the disagreement is larger when the \textit{matter prescription} is used. 

In Fig. \ref{bias_CF_non_linear} we show the results in terms of the other 
bias estimator: $b_{\rm hh}^2(r)=\xi_{\rm hh}(r)/\xi_{\rm mm}(r)$. Again,
we find the Tinker prediction to work well with the \textit{cold dark matter prescription}. 
The bias scale-dependence is also present in the cosmologies with massive
neutrinos when using this definition for the bias. On small scales, the 
correlation function drops quickly for
radii smaller than $r\sim2,3$ $h^{-1}$Mpc. This is due to the finite
size of the dark matter haloes, i.e. the halo exclusion
effect.

\begin{figure}
\begin{center}
\includegraphics[width=1.0\textwidth]{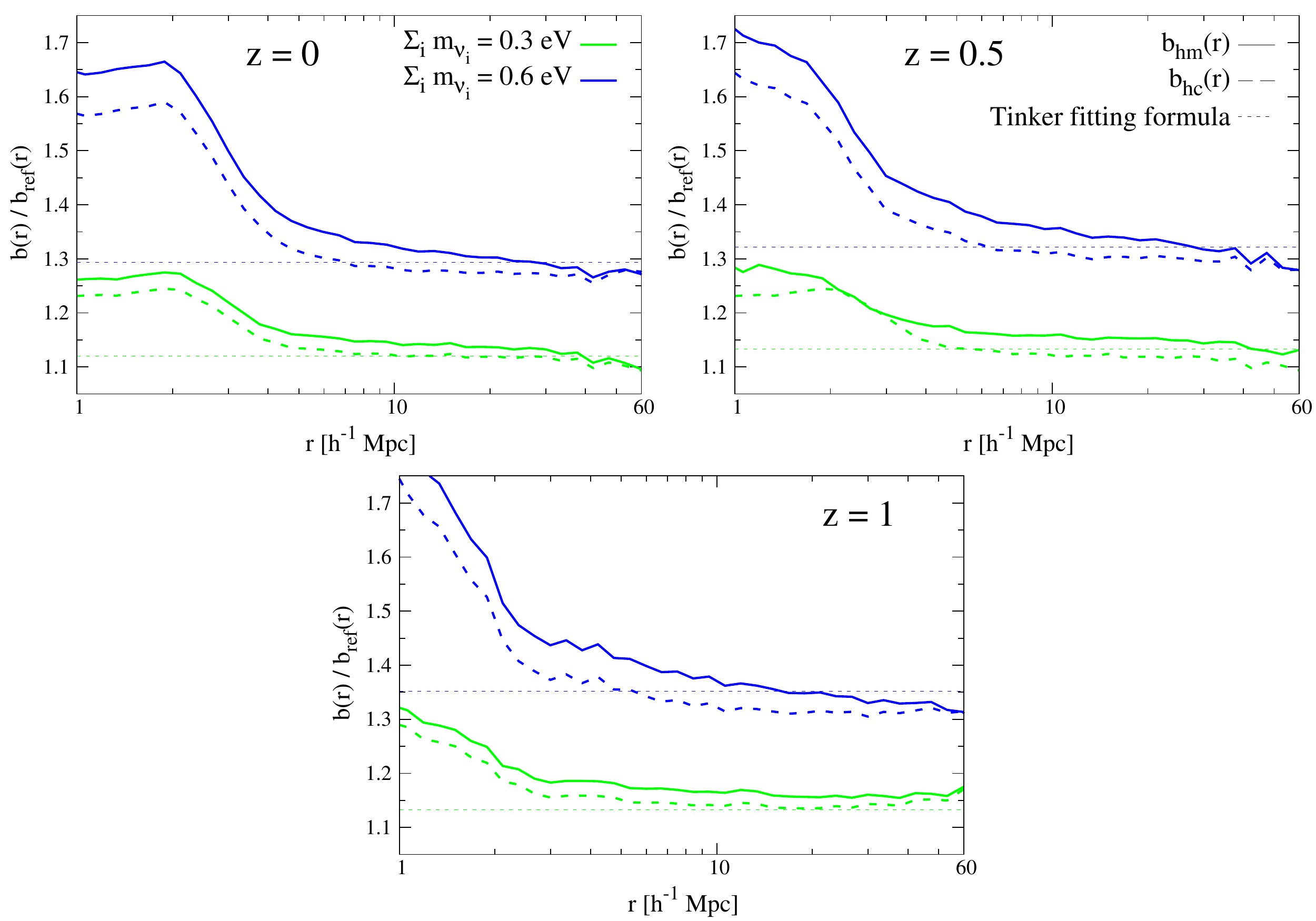}
\end{center}
\caption{With solid lines we show the bias, computed as ratio of the
  haloes-matter cross-correlation function to the matter 2-pt
  autocorrelation function, for the models with \Mnu = 0.3 eV (green
  lines) and \Mnu = 0.6 eV (blue lines) normalized by the bias of the
  model with massless neutrinos. The dashed lines represent the
  results when the bias is computed with respect to the CDM
  distribution, i.e. $b_{\rm hc}(r)=\xi_{\rm hc}(r)/\xi_{\rm
    cdm}(r)$. The prediction of the Tinker fitting formula, using the
  \textit{cold dark matter prescription} for massive neutrinos, is
  shown with dotted lines.}
\label{bias_CF}
\end{figure}

We now check whether the scale-dependent bias we find in the
simulations with massive neutrinos is suppressed if the bias is
computed with respect to the spatial distribution of the CDM, as we
found in the above subsection.
 In Fig. \ref{bias_CF} we plot with solid lines the bias,
computed with respect to the total matter distribution $b_{\rm hm}(r)$,
for the models with \Mnu = 0.3 eV (green) and \Mnu = 0.6 eV (blue), 
normalized by bias of the cosmological model with \Mnu = 0.0 eV. 
The scale-dependent bias is evident on large scales, in particular
for the model with \Mnu = 0.6 eV and at $z=1$. With dashed lines
we show the results of computing the bias with respect to the CDM
distribution, i.e. using $b_{\rm hm}(r)=\xi_{\rm hc}(r)/\xi_{\rm cdm}(r)$, 
with $\xi_{\rm cdm}(r)$ being the correlation function of the CDM
component.
On large scales, the scale-dependence is highly reduced,
resulting in a ratio to the massless case
almost flat. We also show the values of the bias ratio on large scales obtained by
using the Tinker fitting formula along with the cold dark matter
prescription for massive neutrinos. Even though the agreement with our
results is not perfect, in particular for the model with \Mnu = 0.6 eV, it is 
much better than the one we get by using the Tinker fitting formula with 
the matter prescription.

\section{Clustering of galaxies}
\label{sec:HOD}

In this section we populate the haloes of the
different cosmological models with a realistic population of galaxies to study: 1)
whether the distribution of galaxies in cosmological models with
massive neutrinos can reproduce the observed galaxy clustering; and 2) to understand
to what extent the distribution of galaxies within
dark matter haloes, that is on Mpc scales, is affected by massive neutrinos. 
We address
these issues in the framework of halo occupation distribution (HOD)
models that we have adopted to build mock galaxy catalogues.  
These investigations should be regarded 
as a first step towards the more ambitious goal of constraining 
neutrino masses from small scales  galaxy clustering. This section is split 
into 2 subsections. In the subsection 
\ref{HOD_section1} we describe the method we use to construct mock
galaxy catalogues for a particular cosmological model, whereas in the subsection
\ref{HOD_section2} we investigate the impact of neutrino masses into the galaxy 
clustering properties.

\subsection{Construction of mock galaxy catalogues}
\label{HOD_section1}

In this subsection we describe the procedure we use to construct 
mock galaxy catalogues by populating with galaxies the dark matter haloes of a 
particular cosmology using a simple HOD model. 

The HOD  is a 
framework used to link the distribution of galaxies to that of dark matter haloes
\cite{Ma_Fry_2000, Peacock_Smith_2000, Seljak_2000, Scoccimarro_2001,
  Berlind_2002, Kravtsov_2004, Zheng_2004}. Within this framework, the
galaxy bias is completely specified by two ingredients:
1) the probability, $P(N | M)$, that a halo of mass $M$ hosts $N$
galaxies of a given type; 2) the relation between the velocity and spatial 
distribution of the dark matter and the galaxies within haloes.

We now describe the method we use to construct mock galaxy catalogues 
from the snapshots of our N-body simulations.
For a given snapshot of an N-body simulation, we first identify all the dark matter
haloes and extract their properties using the SUBFIND algorithm
\cite{Subfind} (see section \ref{Simulations} for details). The virial 
radius of each halo is defined as the radius at which the mean density within it is
200 times the mean density of the universe. In this work we focus on
the clustering properties of galaxies at low redshift, and thus, we only use
the $z=0$ N-body snapshots.

To populate the dark matter haloes with mock galaxies we follow the scheme
of \cite{Kravtsov_2004,Zheng_2004}, which consists of the following steps:
1) haloes with masses larger than $M_{\rm min}$ contain one central galaxy:
haloes with masses smaller than $M_{\rm min}$ do not host any,
central or satellite, galaxy. Thus, only haloes hosting a central galaxy can host 
satellite galaxies. 2) the number of satellite galaxies
within haloes of mass $M$ follows a Poisson distribution with a mean
given by the power-law $\langle {\rm N}
\rangle_M=\left(M/M_1\right)^\alpha$. We place the central galaxy at
the center of the dark matter halo, and we assume that the
distribution of satellites follows that of the underlying cold dark matter.
We achieve this by setting the satellite galaxies on top of cold dark
matter particles randomly selected within the virial radius of the
halo.

Thus, our HOD model has three free parameters: $M_{\rm min}$, $M_1$
and $\alpha$. Our aim here is to find, for each cosmological model, the values
of those parameters by requiring two conditions: 1) that the galaxy
mock catalogue reproduces the observational galaxy clustering 
measurements; 2) that the number density of the galaxies in the mock matches
the observations. As we will
see below, the HOD parameter values depend on the type of galaxies used to
populate the haloes. We have chosen to populate the dark matter haloes
with galaxies with $M_r-5\log_{10}h<-21$ and $M_r-5\log_{10}h<-21.5$
(where the magnitudes are in the $^{0.1}r$ band), to directly compare
our results with the small scale galaxy clustering data from the 
main galaxy sample of SDSS-II DR7 \cite{Zehavi_2011}.

One of the conditions we require when constructing the mock galaxy catalogues
is that their galaxy number density equals the one obtained from the
observations, in particular, from the luminosity function.
We emphasize that we do not
attempt to create mock galaxy catalogues that reproduce both the
galaxy clustering measurements and the luminosity function of the
galaxies. For the observational luminosity function we use
the Schechter luminosity function with the parameter values
presented in Blanton et al.\footnote{The value of the parameters are:
  $\phi_*=1.49\times10^{-2}~h^3\rm{Mpc}^{-3}$,
  $M_*-5\log_{10}h=-20.44$ and $\alpha=-1.05$. They have been taken
  from \cite{Blanton_2003} } \cite{Blanton_2003}. Thus, the mean
galaxy number density of the mocks containing galaxies brighter than
-21 is equal to $1.11\times10^{-3}~h^3{\rm Mpc}^{-3}$, whereas this
number shrinks to $2.85\times10^{-4}~h^3{\rm Mpc}^{-3}$ for the mocks
containing galaxies with magnitudes smaller than -21.5. 

Chosen a galaxy type, the values of the parameters $M_1$
and $\alpha$ are sampled from the $M_1-\alpha$
plane, whereas the value of $M_{\rm min}$ is chosen by requiring that the
galaxy number density of the mock equals the one obtained from the
luminosity function. Once the values of all the HOD parameters are
selected, we create a mock galaxy catalogue by following the procedure
described above, and compute its 2-pt auto-correlation function
$\xi(r)$. Next, we calculate the projected correlation function of the
mock galaxy catalogue

\begin{equation}
w_p(r_p)=\int_{r_p}^\infty dr \frac{2r\xi(r)}{\sqrt{r^2-r_p^2}}~,
\end{equation}
and compare it with the observational results from \cite{Zehavi_2011}. The comparison 
is quantified using the standard $\chi^2$ statistics
\begin{equation}
\chi^2=\sum\limits_{i,j} \left[w_p(r_{p_i}) - \tilde{w}_p(r_{p_i})\right]C_{ij}^{-1}\left[w_p(r_{p_j}) - \tilde{w}_p(r_{p_j})\right]~,
\label{chi2}
\end{equation}
where $\tilde{w}_p(r_{p_i})$ and $C_{ij}$ are the observational
projected correlation function and its covariance matrix\footnote{The
  covariance matrices have been kindly provided by Idit Zehavi (private
  communication).}. The
$M_{\rm min}-\alpha$ parameter space plane is scanned until the
$\chi^2$ minimum is found. The method used to find the minimum is the
following: we first compute the value of the $\chi^2$ on each of the
points of a coarse-grid where the parameter values are allowed to vary
within a wide interval. Then, we make the grid finer over the
interval in which the smallest values of $\chi^2$ lie. We keep making
the grid finer and finer until the $\chi^2$ minimum is stable.

We note that we are not taking into account the errors associated to
the measurement of the projected correlation function of mock
galaxies. As a consequence, the values of the $\chi^2$ for two mock
galaxy catalogues created with the same values of the HOD parameters
can be different. This is because by changing the random seed we may,
for instance, populate more the outer regions of the dark matter
haloes, and this will affect both the shape and the amplitude of the
projected mock galaxy correlation function. 
 
The 2-pt autocorrelation function of the mock galaxy catalogues
is computed in real-space using
the Landy-Szalay estimator \cite{Landy-Szalay_93}, in the same way as
explained in section \ref{subsec:correlation_function}. The random catalogue is
constructed by filling the simulation box with four hundred thousand
particles, randomly distributed within the interior volume. The number
of points in the random catalogue is between two and ten times the
number of galaxies in our mock catalogues. We have explicitly checked
that for radii larger than $\sim 0.3 ~h^{-1}$Mpc our results, in terms
of the autocorrelation function, do not change if we increase the
number of points in the random catalogue. However, at radii smaller
than $\sim 0.3 ~h^{-1}$Mpc our results are sensitive to the number of
random points and to the binning used to compute the autocorrelation
function. Moreover, on those small scales the SDSS galaxy clustering 
data may be affected by systematic errors arising from fiber collision.
Therefore, we choose not to include the points with
$r_p<0.3~h^{-1}$Mpc in our analysis\footnote{Due to computational
  limitations, we are unable to perform the analysis described here
  with a random catalogue containing a number of points much larger
  than 400000.}. We have also excluded from our analysis the noisy bins
at $r_p>40~h^{-1}$Mpc. The 2-pt autocorrelation function has been
computed in 60 logarithmically equally spaced bins between $r=0.1~h^{-1}$Mpc
and $r=75~h^{-1}$Mpc. We have checked that our results are converged
against the number of intervals and the maximum radius used to count
pairs\footnote{Our tests are performed over the projected
  correlation function.}.

\begin{figure}
\begin{center}
\includegraphics[width=1\textwidth]{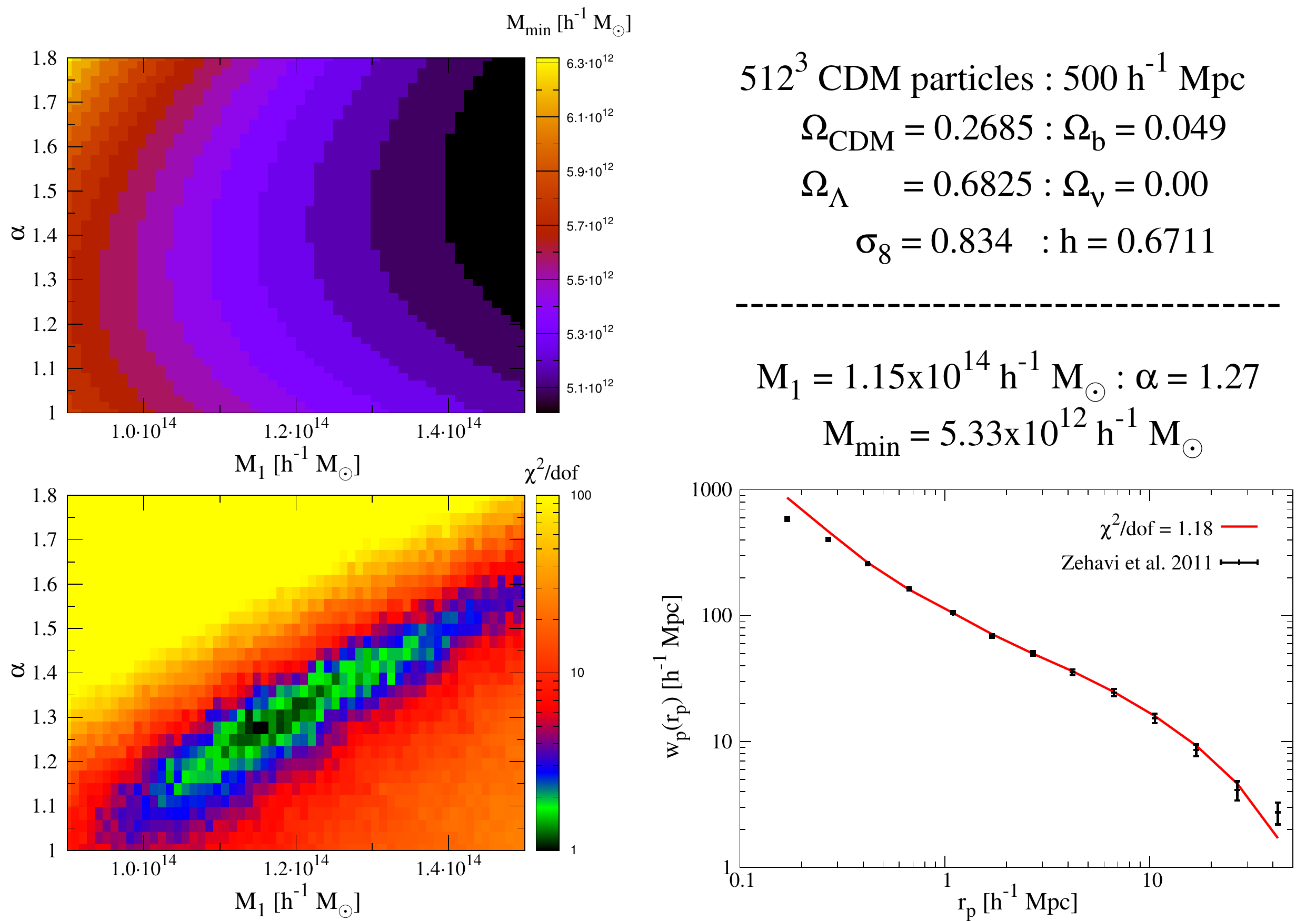}
\end{center}
\caption{An example of the HOD analysis, performed in this case on top
  of the L0-Planck simulation. The cosmological parameters of that
  simulation are reported in the \textit{upper-right} panel. We
  construct mock galaxy catalogues, for galaxies with
  $M_r-5\log_{10}h<-21$, requiring that their number density to be
  equal to this obtained from the luminosity function of Blanton et
  al. \cite{Blanton_2003}. The value of the parameter $M_{\rm min}$ is
  fixed by demanding the former condition. In the \textit{upper-left}
  panel we show the value of $M_{\rm min}$ as a function of the
  parameters $\alpha$ and $M_1$. For each point in the HOD parameter
  space we create a mock galaxy catalogue and compute its projected
  correlation function, comparing it with observations by using the
  $\chi^2$ statistics. In the \textit{bottom-left} panel we plot the
  value of $\chi^2$, normalized to the number of degrees of freedom,
  as function of $\alpha$ and $M_1$. The values of the HOD parameters
  that better reproduce observations for this cosmology are shown at
  the bottom of the \textit{upper-right} panel. The projected
  correlation function for those HOD parameters is shown in the
  \textit{bottom-left} panel, together with the observational
  measurements. Note that when comparing with observations we only use
  the $w_p$ points with $r>0.3$ $h^{-1}$ Mpc and $r<30$ $h^{-1}$ Mpc.}
\label{HOD}
\end{figure}
 
We create mock galaxy catalogues, using the procedure described
above, for cosmologies with massless and massive neutrinos to
investigate how well our mock galaxy catalogues can reproduce 
the SDSS small scale galaxy clustering measurements, and to studying how galaxies
should populate the dark matter haloes in the different cosmological models. 
As an example, we show in Fig. \ref{HOD} the results of
populating the dark matter haloes, at $z=0$, of the simulation L0-Planck with galaxies with
magnitudes $M_r-5\log_{10}h<-21$. The values of the cosmological
parameters of the simulation, together with the values of the HOD parameters that
better reproduce the measured projected correlation function are shown
in the upper-right panel. The upper-left panel shows the value of the
parameter $M_{\rm min}$ as a function of $M_1$ and $\alpha$. We find that the function
$M_{min}=M_{min}(M_1,\alpha)$ displays a maximum when moving along
lines with $M_1$ constant. In order to understand that feature, we
should take into account two competing effects. The mean number
density of satellite galaxies is given by:
\begin{equation}
\langle N \rangle_{\rm satellites} = \int_{M_{\rm min}}^{\infty}\left(\frac{M}{M_1}\right)^\alpha n(M,z) dM~,
\end{equation}
where $n(M,z)dM$ is the number density of haloes with masses between
$M$ and $M+dM$. Whereas the number of satellite galaxies residing in
haloes with masses smaller than $M_1$ decrease with increasing
$\alpha$, the number of satellites galaxies within haloes of
masses larger than $M_1$ increases with $\alpha$. For low values of
$\alpha$, the total number of satellites is dominated by those living
in haloes with masses higher than $M_1$, and therefore, as $\alpha$
increases, the total number of satellites increases. This implies that
the number of central galaxies should shrink (to keep the same galaxy
number density), i.e. the value of $M_{\rm min}$ should increase with
$\alpha$. If the value of $\alpha$ is large, the satellite population
will be dominated by galaxies residing in haloes with masses smaller
than $M_1$. In this situation, a larger value of $\alpha$ will further
shrink the number of satellites. Therefore, the number of central
galaxies should increase, which implies that the value for $M_{\rm min}$
should decrease with increasing $\alpha$.  

On the bottom-left panel we plot the value of the $\chi^2$, normalized
to the number of degrees of freedom, $\chi^2$/d.o.f., as a function of
the values of the HOD parameters. As stated above, the comparison between
the clustering of our mock galaxy catalogues and the SDSS data
is performed for points in the projected correlation function with 
radii larger than 0.3 $h^{-1}$Mpc and
smaller than 40 $h^{-1}$ Mpc by using the full covariance matrix. We
find a strong correlation between the parameters $\alpha$ and
$M_1$. We note that because we are not taking into account the errors
in the measurement of the mock galaxy correlation function, the values
of the $\chi^2$ not only depend on the HOD parameters but also on the 
mock catalog realization, i.e. on the set of random numbers we use to
populate the dark matter haloes with galaxies. This is clearly reflected in the
$\chi^2=\chi^2(\alpha,M_1)$ plot: the $\chi^2$ function is not a
smooth function of the parameters $\alpha$ and $M_1$, but the scatter
induced by the particular realization used to create the mock galaxy
catalogue appears as bins with colors slightly displaced to the color
they would have if the mean of several realizations would be use. 

Finally, the projected correlation function of the mock galaxy catalogue
constructed with values of the HOD parameters that 
minimize the $\chi^2$ is plotted on the bottom-right
panel. The black points represent the observational measurements, 
taken from Zehavi et al. \cite{Zehavi_2011}, where the error bars 
correspond to the square root of
the covariance matrix diagonal terms. We find that with this simple
HOD model we can reproduce pretty well the projected correlation function for
galaxies with magnitudes brighter than -21, obtaining a
minimum value of $\chi^2$/d.o.f. equal to 1.18. We quote the values
of the HOD parameters that minimize the $\chi^2$ on table \ref{HOD_parameters}.

\begin{figure}
\begin{center}
\includegraphics[width=1.0\textwidth]{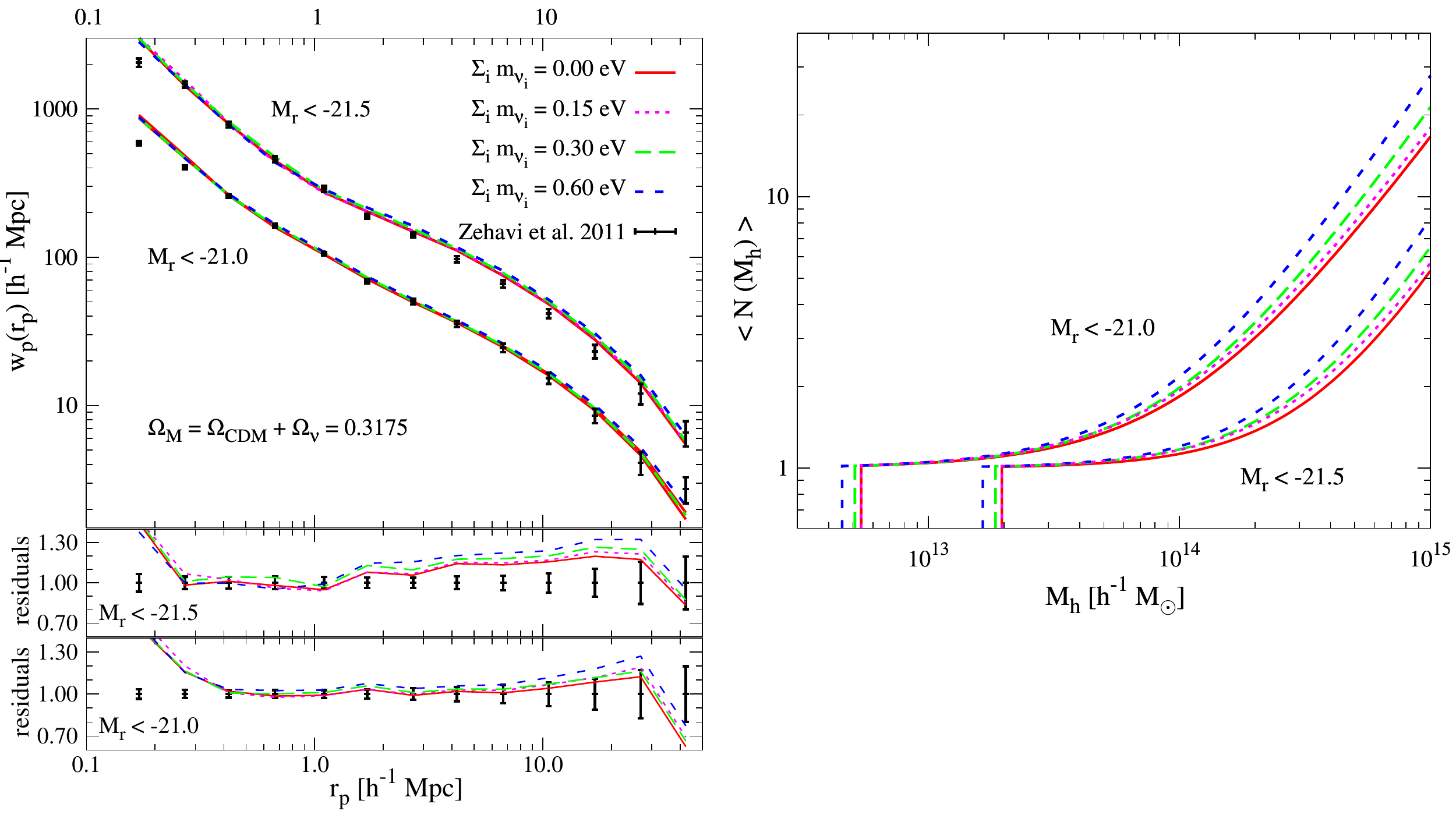}
\end{center}
\caption{In the upper-right panel we show with black points the $w_p$
  measurements for galaxies with $M_r-5\log_{10}h<-21$ (bottom points)
  and $-21.5$ (top points) from Zehavi et al. \cite{Zehavi_2011}. The
  error bars represent the diagonal terms of the covariance
  matrix. The projected correlation function of the mock galaxy
  catalogues created with HOD parameter values that better reproduce
  those observations are shown for three different cosmologies with
  different sums of the neutrino masses: 0 eV (solid red), 0.3 eV
  (long dashed green) and 0.6 eV (short dashed blue). For clarity, the
  observational measurements and our results for the galaxies with
  magnitudes brighter than -21.5 have been displaced vertically by a
  factor of 2. The relative differences between those and observations
  are show in the middle panel (for galaxies with
  $M_r-5\log_{10}h<-21.5$) and bottom panel (for galaxies with
  $M_r-5\log_{10}h<-21$). The right panel shows the mean number of
  galaxies as a function of the mass of the dark matter halo hosting
  them for the three different cosmologies.}
\label{PTPCF_planck}
\end{figure}

\subsection{HOD: galaxy clustering properties}
\label{HOD_section2}

In the above section we have explained the method we use to construct
the mock galaxy catalogues for a particular cosmological model. In this
subsection we investigate how well our mock galaxy catalogues can reproduce
the observational clustering measurements, and how the distribution of galaxies
within dark matter haloes is impacted by the masses of neutrinos.

We first investigate how well we can reproduce the observed projected
correlation function for cosmologies with massive neutrinos. We construct
mock galaxy catalogues, containing galaxies with magnitudes
brighter than -21, for those cosmological models by repeating the above
procedure on top of the simulations with \Mnu = 0.15 eV (L15-Planck),
\Mnu = 0.30 eV (L3-Planck) and \Mnu = 0.60 eV (L6-Planck). We note 
that those simulations share the
value of the parameters $\Omega_{\rm cdm}$ and $A_s$ with the
simulation with massless neutrinos used above (L0-Planck). The
projected correlation functions of the mock galaxy catalogues created
with the HOD parameters that better reproduce the clustering
observations are shown in the upper-left panel of
Fig. \ref{PTPCF_planck}. In the bottom-left panel of that figure we
show the relative difference between observations and our mock
catalogues. We find that the models with massive neutrino can also
reproduce pretty well the observational measurements: $\chi^2$/d.o.f. =
1.28, 1.33 and 1.35 for the cosmological models with neutrino masses
equal to \Mnu = 0.15, 0.30 and 0.60 eV, respectively. However, we find 
that HOD models over-predict galaxy correlation at large separations, 
where the 2-halo term dominates.

We repeat the whole analysis, for the same above cosmological
models, by creating mock galaxy catalogues containing galaxies with
magnitudes $M_r-5\log_{10}h<-21.5$. The results are also shown in
Fig. \ref{PTPCF_planck}. For this case, we find that the the galaxy
mocks catalogues do not reproduce that well the observations:
$\chi^2$/d.o.f. = 2.25, 2.73, 3.25 and 4.88 for the models with \Mnu = 0.00,
0.15, 0.30 and 0.60 eV respectively. As for the mocks with galaxies
brighter than -21, we find that the disagreement between our results
and observations increases with the neutrino masses. The disagreement
arises primarily on large scales, where the 2-halo term dominates. The
right panel of Fig. \ref{PTPCF_planck} shows the average number of
galaxies hosted by dark matter haloes as a function of the halo mass,
obtained directly from the values of the HOD parameters that best 
reproduce the observations. For galaxies with luminosities above a certain
threshold, we find that the most massive haloes should contain more
galaxies in cosmologies with massive neutrinos than in cosmologies
with massless neutrinos. Moreover, those galaxies can be hosted by
haloes of lower mass for the massive neutrino cosmologies.  

\begin{figure}
\begin{center}
\includegraphics[width=1.0\textwidth]{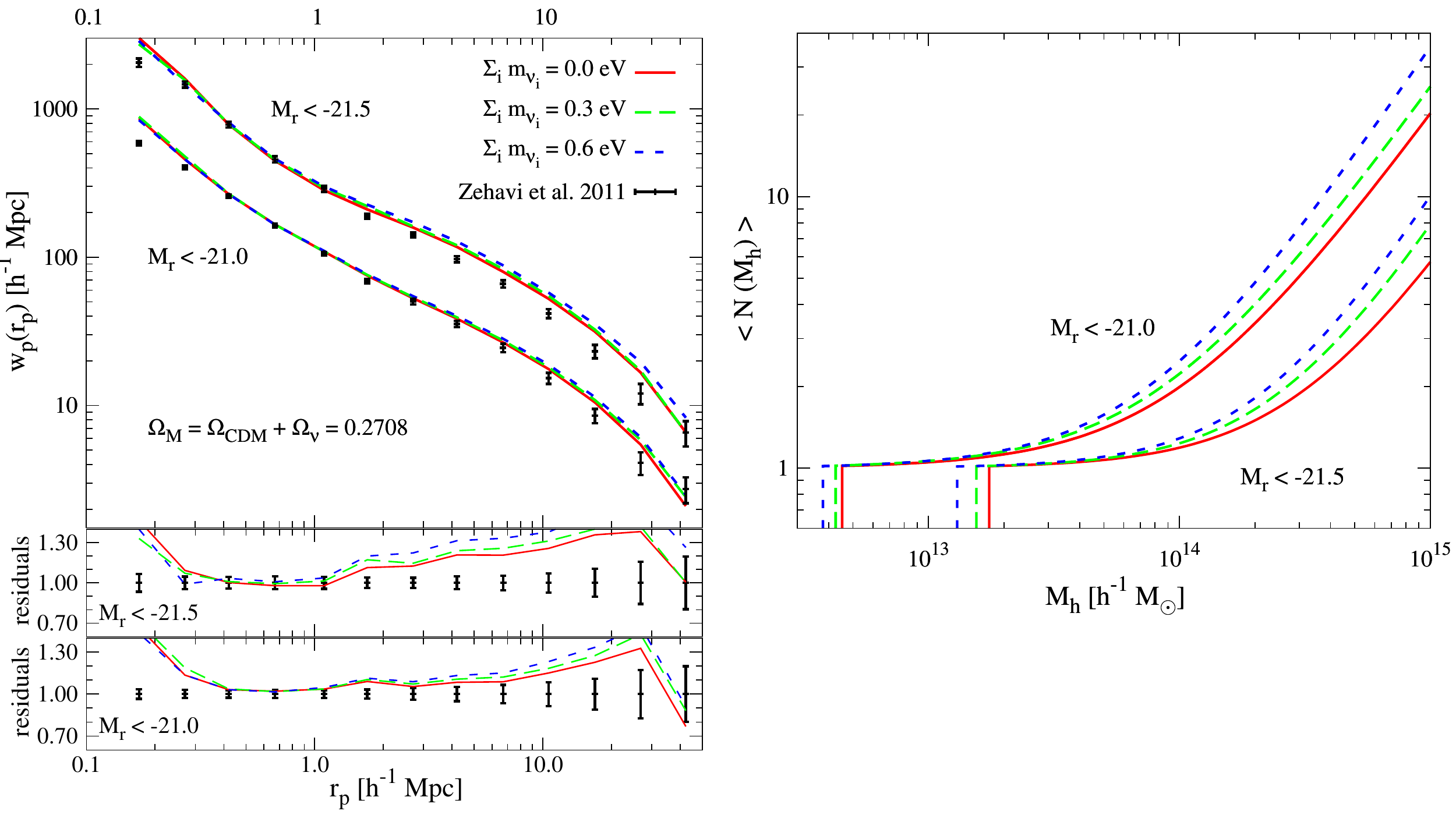}
\end{center}
\caption{Same as Fig. \ref{PTPCF_planck} but with cosmological models
  with $\Omega_{\rm m}=0.2708$ instead of $\Omega_{\rm m}=0.3175$.}
\label{PTPCF_default}
\end{figure}

Now we focus our analysis into the impact of the cosmological parameter 
$\Omega_{\rm m}$ on the clustering properties of our mock galaxy 
catalogues. We have constructed mock galaxy
catalogues using the $z=0$ snapshots of the simulations 
L0, L3 and L6, which basically differ
from the previous ones in the value of $\Omega_{\rm m}$ (see table
\ref{tab_sims}). The results are
shown in Fig. \ref{PTPCF_default}. Using a higher value for $\Omega_{\rm m}$
 further increases the mismatch between HOD model and observations
that, as in the previous tests, increases with scale. For the mocks containing galaxies
with $M_r-5\log_{10}h<-21$ the values of the minimum $\chi^2$/d.o.f. are
1.85, 2.39 and 2.74 for the cosmologies with \Mnu = 0.0, 0.3 and 0.6 eV
respectively. The disagreement is even worst for the mock catalogues hosting
galaxies with $M_r-5\log_{10}h<-21.5$: $\chi^2$/d.o.f. = 3.88, 5.44 and
8.13 for the models with \Mnu = 0.0, 0.3 and 0.6 eV respectively. Again,
we find that the disagreement between our mock catalogues and the
observational data increases with the neutrino masses, arising mainly
on large scales. For each cosmological model and for each mock galaxy
catalogue we list on table \ref{HOD_parameters} the values of the
HOD parameters that minimize the $\chi^2$/d.o.f. together with the value of it.

\begin{figure}
\begin{center}
\includegraphics[width=1.0\textwidth]{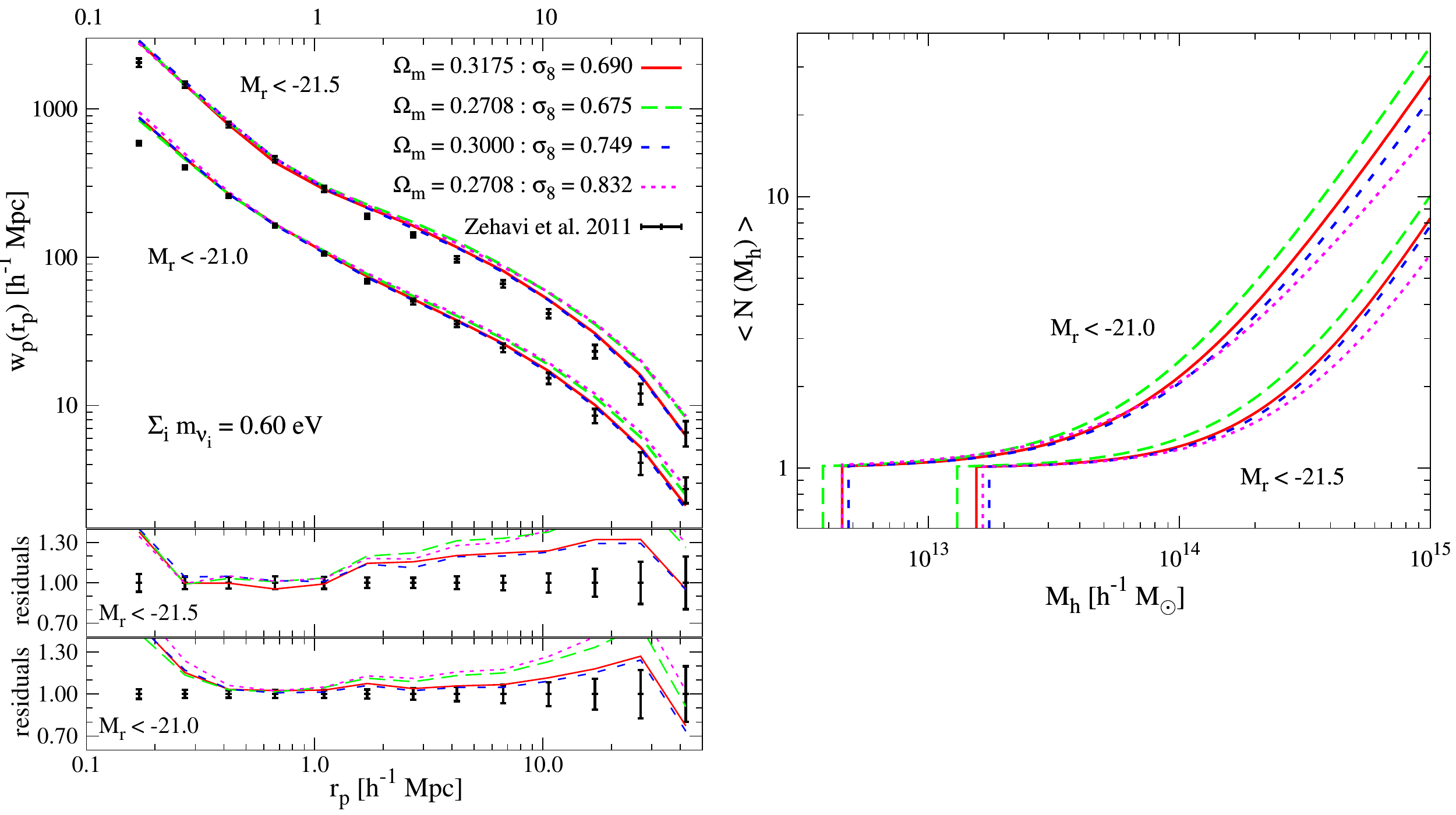}
\end{center}
\caption{Same as Fig. \ref{PTPCF_planck} but with cosmological models
  with neutrinos with masses equal to \Mnu = 0.6 eV. The mock galaxy
  catalogues have been constructed over the simulations L6-Planck
  (red), L6 (green), L6-2 (blue) and L6-1 (magenta). The values of the
  parameters $\Omega_{\rm m}$ and $\sigma_8$ are shown on the
  top-right corner of the upper-left panel for the different
  cosmological models.}
\label{PTPCF_06}
\end{figure}

Next, we study which is the influence of the parameter $\Omega_\nu$
on the clustering properties of our mock catalogues. In particular, we
would like to know if the main effect in the galaxy clustering
properties arise from $\Omega_\nu$, or whether it is mainly related to
the values of other cosmological parameters such as $\Omega_{\rm cdm}$
or $\sigma_8$. To answer this question we have created mock galaxy
catalogues using the simulations L6-Planck, L6, L6-1 and L6-2. All
those simulations represent cosmological models with neutrinos of
masses \Mnu = 0.60 eV. The differences among them are in the value of
the cosmological parameters $\Omega_{\rm m}$ and $\sigma_8$. The value
of the couple $(\Omega_{\rm m},\sigma_8)$ is (0.3175, 0.69), (0.2708,
0.675), (0.2708, 0.832) and (0.30, 0.749) for the simulations
L6-Planck, L6, L6-1 and L6-2 respectively. The projected correlation 
function of the HOD galaxies that best fits observations is shown in 
Fig. \ref{PTPCF_06}, and the HOD parameters together the
value of the $\chi^2$/d.o.f. are listed in table \ref{HOD_parameters}. We
find that the simulations in which our HOD model reproduces better the
observed galaxy clustering are L6-Planck and L6-2. Those simulations
have values of the parameter $\Omega_{\rm cdm}$ higher than the
simulations L6 and L6-1. The impact of the parameter $\sigma_8$ seems
to be less important. The simulation L6-1 has a value of
$\sigma_8=0.832$ significantly larger than all the other simulations,
but we find that values of the $\chi^2$/d.o.f. equal to 3.81 and 6.71 for
galaxies with magnitudes brighter than -21 and -21.5 respectively, in
comparison with the simulations L6-Planck and L6-2, for which we find
$\chi^2$/d.o.f. = 1.35 and 4.88 (for L6-Planck) and $\chi^2$/d.o.f. = 1.32
and 3.42 (for L6-2). However, the value of $\sigma_8$ may be the
reason why our mock galaxy catalogue reproduces better the
measurements for galaxies with magnitudes smaller than -21.5 in the
case of the simulation L6-2 than in the simulation L6-Planck. Overall,
we find that our mock galaxy catalogues reproduce pretty well the
projected correlation function on small scales (the 1-halo term) for
all the cosmological models.

It can be argued that the values of the parameters $\Omega_\Lambda$, $h$ or
$n_s$ also play an important role on our results. However since the models that 
best fit the data, all models with the suffix -Planck plus model L6-2,
use a similar $\Omega_{\rm cdm}$ while the other parameters span a wide
range of values suggests that galaxy clustering is mostly sensitive to the value of the
cosmological parameter $\Omega_{\rm cdm}$.

We have checked the robustness of our results by running a N-body
simulation with the same mass resolution as the previous simulations
but with a box size twice as large. This simulation, labeled as
H0-Planck-LV on table \ref{tab_sims} has the same cosmological
parameters as the simulation L0-Planck, although the former has a box
size equal to 1000 $h^{-1}$ Mpc whereas the size of the latter is of
500 $h^{-1}$ Mpc. The resolution in both simulations is the same since
H0-Planck-LV contains $1024^3$ CDM particles (to compare with the
$512^3$ of L0-Planck). We have computed the values of the HOD
parameters for which the mock galaxy catalogues best reproduce
the projected correlation function of
galaxies with $M_r-5\log_{10}h<-21.5$\footnote{Note that for galaxies
  with $M_r-5\log_{10}h<-21$ the mock catalogues would contain more
  than 1 million of galaxies, making the analysis presented here too
  expensive from the computational point of view.}. We find that the
best fit HOD parameters 
are slightly different to those obtained for the simulation
L0-Planck. However, the minimum value of 
$\chi^2$/d.o.f. remains the
same, pointing out that the discrepancies arise from the 2-halo term,
which, in our simple HOD implementation, is mainly determined by 
the cosmological model. We emphasize that the purpose of this work is not to determine
the exact values of HOD parameters that reproduce the observations, but to
study the impact of massive neutrinos on the galaxy clustering properties. Thus,
our main aim here is to investigate the relative differences among the
different models.

The pretty large disagreement between the projected correlation
function of our $M_r-5\log_{10}h<-21.5$ mock galaxy catalogues and the
observed one may be due to the simplistic HOD model we have used
to populate the dark matter haloes. Zehavi et al. 2004
\cite{Zehavi_2004} found that the projected correlation function of
the brightest galaxies is better reproduced if a different HOD
model is used. The large values of the $\chi^2$ may also be due to an
underestimation of the covariance matrix, as claimed also by
\cite{Zehavi_2011}.

\begin{table}
\begin{center}
\begin{tabular}{|c|c|c|c|c|c|c|c|c|}
\hline
$M_r-5\log_{10}h$ & Simulation & $\Omega_{\rm m}$ & $\sigma_8 $&\Mnu & $M_1$ & $\alpha$ & $M_{\rm min}$ & $\chi^2$/d.o.f.\\
& name & & (z=0) & (eV) & ($h^{-1}\rm{M}_\odot$) & & ($h^{-1}\rm{M}_\odot$) &\\
\hline
\hline 

-21.0 & L0-Planck & 0.3175 & 0.834 & 0.00 & $1.15\times10^{14}$ & 1.27 & $5.33\times10^{12}$ & 1.18\\
\hline
-21.0 & L15-Planck & 0.3175 & 0.799 & 0.15 & $1.06\times10^{14}$ & 1.26 & $5.19\times10^{12}$ & 1.28\\
\hline
-21.0 & L3-Planck & 0.3175 & 0.761 & 0.30 & $1.02\times10^{14}$ & 1.32 & $4.91\times10^{12}$ & 1.33\\
\hline
-21.0 & L6-Planck & 0.3175 & 0.690 & 0.60 & $8.90\times10^{13}$ & 1.36 & $4.47\times10^{12}$ & 1.35\\
\hline
-21.0 & L0 & 0.2708 & 0.832 & 0.00 & $1.01\times10^{14}$ & 1.29 & $4.48\times10^{12}$ & 1.85\\
\hline
-21.0 & L3 & 0.2708 & 0.752 & 0.30 & $8.56\times10^{13}$ & 1.30 & $4.18\times10^{12}$ & 2.39\\
\hline
-21.0 & L6 & 0.2708 & 0.675 & 0.60 & $7.52\times10^{13}$ & 1.37 & $3.73\times10^{12}$ & 2.74\\
\hline
-21.0 & L6-1 & 0.2708 & 0.832 & 0.60 & $9.39\times10^{13}$ & 1.18 & $4.30\times10^{12}$ & 3.81\\
\hline
-21.0 & L6-2 & 0.3000 & 0.749 & 0.60 & $9.59\times10^{13}$ & 1.32 & $4.67\times10^{12}$ & 1.32\\
\hline \hline

-21.5 & L0-Planck & 0.3175 & 0.834 & 0.00 & $3.84\times10^{14}$ & 1.53 & $1.95\times10^{13}$ & 2.25\\
\hline
-21.5 & L15-Planck & 0.3175 & 0.799 & 0.15 & $3.45\times10^{14}$ & 1.45 & $1.88\times10^{13}$ & 2.73\\
\hline
-21.5 & L3-Planck & 0.3175 & 0.761 & 0.30 & $3.23\times10^{14}$ & 1.51 & $1.77\times10^{13}$ & 3.25\\ 
\hline
-21.5 & L6-Planck & 0.3175 & 0.690 & 0.60 & $2.79\times10^{14}$ & 1.56 & $1.55\times10^{13}$ & 4.88\\
\hline
-21.5 & L0 & 0.2708 & 0.832 & 0.00 & $3.29\times10^{14}$ & 1.40 & $1.66\times10^{13}$ & 3.88\\
\hline
-21.5 & L3 & 0.2708 & 0.752 & 0.30 & $2.69\times10^{14}$ & 1.47 & $1.48\times10^{13}$ & 5.44\\ 
\hline
-21.5 & L6 & 0.2708 & 0.675 & 0.60 & $2.30\times10^{14}$ & 1.50 & $1.28\times10^{13}$ & 8.13\\
\hline
-21.5 & L6-1 & 0.2708 & 0.832 & 0.60 & $3.32\times10^{14}$ & 1.48 & $1.58\times10^{13}$ & 6.71\\ 
\hline
-21.5 & L6-2 & 0.3000 & 0.749 & 0.60 & $2.95\times10^{14}$ & 1.57 & $1.65\times10^{13}$ & 3.42\\
\hline

\end{tabular}
\end{center} 
\caption{Values of the HOD parameters, for each cosmological model and
  for with magnitudes than a given threshold, that best fit the
  observational measurements.}
\label{HOD_parameters}
\end{table}

Our results suggest that with the simple HOD modeling we have used in
this paper we can construct mock galaxy catalogues that reproduce the
SDSS small scale galaxy clustering data, for radii
smaller than $\sim1.5~h^{-1}$Mpc, for all the cosmological models,
independently of the neutrino masses or the values of $\Omega_{\rm
  m}$ and $\sigma_8$. However, we find that for radii larger than $\sim2~h^{-1}$Mpc,
the shape and amplitude of the projected correlation function of our
mock catalogues is mainly determined by the value of the cosmological
parameter $\Omega_{\rm cdm}$. Since the cosmological models we have
used in this paper share the value of the parameter $\Omega_{\rm
  m}=\Omega_{\rm cdm}+\Omega_\nu$, we find that the largest the
neutrino masses the largest the disagreement on large scales, but this
is just a consequence of the different value of the parameter
$\Omega_{\rm cdm}$ among the models.

Finally, we quantify the dependence of the values of the HOD
parameters with the neutrino masses. Table \ref{HOD_parameters}
shows the values of the HOD parameters, for each mock catalogue we have
created, that better reproduce the SDSS galaxy clustering measurements.
We find that the parameter $\alpha$ exhibits a very weak
dependence on the masses of the neutrinos and with cosmology.
However, when comparing models with the same values of the parameters
$\Omega_{\rm m}$ and $A_s$, we find that the general trend indicates
that this parameters increases with $\Sigma_i m_{\nu_i}$. On the other hand, the
values of the parameters $M_1$ and $M_{\rm min}$ are strongly
dependent on the neutrino masses. By comparing the values of those
parameters among cosmological models that share the values of
$\Omega_{\rm m}$ and $A_s$ (L6-Planck, L3-Planck, L15-Planck and
L0-Planck on one hand, and L6, L3 and L0 on the other hand) we find
that a simple relation of the form $\log (M_1)=a+b$\Mnu and $\log
(M_{\rm min})=a+b$\Mnu reproduces pretty well the dependence of those
parameters with the neutrino masses. 
In table \ref{HOD_parameters_Mnu} we show the values of the parameters 
$a$, $b$, $c$ and $d$ for the two different set of cosmological models and for 
the two types of mock galaxy catalogues.

\begin{table}
\begin{center}
\begin{tabular}{|c|c|c|}

\hline
$M_r-5\log_{10}h$ & $\Omega_{\rm m}=0.3175$ & $\Omega_{\rm m}=0.2708$ \\
 & (L0-Planck, L15-Planck, L3-Planck, L6-Planck) & (L0, L3, L6) \\
\hline \hline
-21.0 & $\log(M_1)=32.37-0.416$\Mnu & $\log(M_1)=32.24-0.492$\Mnu \\
 & $\log(M_{\rm min})=29.31-0.302$\Mnu & $\log(M_{\rm min})=29.14-0.305$\Mnu \\
\hline
-21.5 & $\log(M_1)=33.57-0.519$\Mnu & $\log(M_1)=33.42-0.597$\Mnu \\
 & $\log(M_{\rm min})=30.61-0.391$\Mnu & $\log(M_{\rm min})=30.45-0.433$\Mnu \\
\hline

\end{tabular}
\end{center} 
\caption{Dependence of the HOD parameters $M_1$ and $M_{\rm min}$ with the masses of the neutrinos for cosmological models with $\Omega_{\rm m}=0.3175$ (simulations L0-Planck, L15-Planck, L3-Planck and L6-Planck) and with $\Omega_{\rm m}=0.2708$ (simulations L0, L3 and L6). The fits are shown for mock galaxy catalogues with galaxies with magnitudes brighter than -21 and -21.5.}
\label{HOD_parameters_Mnu}
\end{table}

\section{Summary and conclusions}
\label{Conclusions}

We have run a large suite of 80 N-body simulations containing CDM and
neutrino particles to study the impact of massive neutrinos on the
spatial distribution of dark matter haloes and galaxies. 
Our simulations span a total of 12 different cosmological
models, with box sizes between 500 and 1000 $h^{-1}$Mpc. For 6 of
those cosmological model we have run 8 independent realizations, in
order to make our conclusions more robust.

We have investigated the bias between the spatial distribution of dark
matter haloes and that of the underlying matter with two different
techniques: the power spectrum and the correlation function. With both
techniques, we find that on large scales and for the simulations with massless
neutrinos, the bias becomes scale-independent, in agreement with theoretical
expectations. However, in the simulations with massive neutrinos we
find a weak, scale-dependent bias on large scales. This is true as long as 
the total matter power spectrum, including contributions from both neutrinos 
and CDM, is used in the bias estimator. If, instead, one consider the CDM power 
spectrum, then the scale dependence becomes much weaker. In
our companion paper \cite{Castorina}, we study in detail this point,
showing that the bias becomes universal only if it is computed with
respect to the distribution of CDM.

We have also adressed the $\Omega_\nu-\sigma_8$ degeneracy on the
halo-matter bias. We find that bias is little sensitive to \Mnu if the value of the parameter
$\sigma_8$ is kept fixed. However, in the cosmologies
with massive neutrinos the bias does show some scale dependence. 
A nearly perfect degeneracy will hold between models sharing
the value of $\sigma_8$, if the bias is computed with respect to the
CDM linear power spectrum.

For cosmologies with massive neutrinos we find that the value of the
bias on large scales is reasonably well described by the fitting
formula of Tinker \cite{Tinker_2010}, once the \textit{cold dark
  matter prescription} for massive neutrinos is used. This
prescription relies on the use of the CDM linear matter power spectrum
when computing the value of $\sigma(M)$ (or equivalently
$\nu=\delta_c/\sigma(M)$), instead of the total matter linear power
spectrum (\textit{matter prescription}), and in replacing $\rho_{\rm
  m}$ with $\rho_{\rm cdm}$.

By populating the dark matter haloes with realistic galaxies we have
constructed mock galaxy catalogues for a large variety of cosmological
models, and we have used these simulations to investigate the impact
of neutrino masses on galaxy clustering properties. We have used a
simple HOD model to populate the haloes with galaxies, whose
magnitudes, in the $^{0.1}r$ band, are smaller than -21 and -21.5. We
have constructed our mock catalogues by requiring that: 1) the number
density of the galaxies in the mocks is equal to the one obtained from
the luminosity function of Blanton et al. \cite{Blanton_2003}; 2)
the clustering properties of the galaxies in the mocks mimic, as close
as possible, the small scales galaxy clustering data of the SDSS
collaboration \cite{Zehavi_2011}. For each cosmological model (N-body
simulation) we tuned the values of the HOD parameters to achieve the
requirements.

Overall, our mock galaxy catalogues can reproduce very well the
projected correlation function of Zehavi et al. on small scales for
all the cosmological models (the so-called 1-halo term). On large scales, 
however, where the projected correlation function of our mocks is determined 
mainly by the cosmological parameter $\Omega_{\rm cdm}$ models 
do not match observations that well. Our mock catalogues can reproduce the
observed projected correlation function of galaxies with
$M_r-5\log_{10}h<-21$ with a $\chi^2$/d.o.f. $\sim$ 1.2-1.4, almost
independently of the value of the sum of the neutrino masses, when
using cosmological models with $\Omega_{\rm m}=0.3175$ (simulations
L6-Planck, L3-Planck and L0-Planck). The values of the $\chi^2$/d.o.f.
are more sensitive to the neutrino masses if a cosmology with a lower
value of $\Omega_{\rm m}=0.2708$ is used: $\chi^2$/d.o.f. $\sim$
1.9-2.7 for the simulations L6, L3 and L0. We also find that the 
value of the cosmological parameter
$\sigma_8$, for this HOD modeling, does not play an important role for
the clustering of galaxies with magnitudes brighter than -21, but with
larger values of $\sigma_8$ we can reproduce better the clustering
measurements for galaxies with magnitudes smaller than -21.5.

Our mock catalogues cannot reproduce the projected correlation
function of galaxies with magnitudes $M_r-5\log_{10}h<-21.5$ well 
as indicated by the large $\chi^2$/d.o.f. values we get: 
from $\sim2-5$ for the cosmological models with $\Omega_{\rm m}=0.3175$ 
to $\sim3-8$ for the cosmological models with $\Omega_{\rm m}=0.2708$. 
We believe that this
may be due to the simplistic HOD model used to populate haloes with
those galaxies. In \cite{Zehavi_2004} authors found than a more
complex HOD parametrization could better reproduce the observational
measurement for the more luminous galaxies. 

We emphasize that the tensions we find, in the projected correlation function, on the 2-halo term
between the observational measurements and our mock galaxy catalogues can be also
alleviated by introducing some scatter in the galaxy number density from the luminosity function,
i.e. by relaxing the condition we require to our mocks: that the galaxy number density of our
mocks has to match the galaxy number density obtained from the luminosity function.
Thus, we believe that the use of a more complex HOD model 
\cite{2013MNRAS.430..725V, 2013MNRAS.430..747M, 2013MNRAS.430..767C} 
may help to reduce the tensions between our mocks and the observational measurements.

We have compared cosmological models that share the values of
$\Omega_{\rm m}$ and $\sigma_8$ but differ on $\Sigma_i m_{\nu_i}$. Our results
indicate that the dark matter haloes in the cosmologies with the most
massive neutrinos have to be more densely populated to match observations.
Moreover, in a massive neutrino cosmology the same galaxy population is hosted 
in less massive haloes that in the massless neutrino cosmology. This 
is however just a reflection of the impact of
neutrino masses on the halo mass function.

Thus, we conclude that it is unlikely that the neutrino masses can
be constrained by using small-scale galaxy clustering measurements 
alone. However, by combining galaxy clustering data with CMB 
measurements, that would fix the total amount of cold dark matter, 
tight bounds on the neutrino masses can be set by exploiting the small 
scale clustering regime of galaxy survey.

\section*{Acknowledgements}
We thank Idit Zehavi for having kindly provided us with the covariance
matrices of the $w_p(r_p)$ measurements. We also thank Ravi Sheth for 
useful comments. FVN would like to thank
Simeon Bird for his help with the N-body simulations run for this
paper. Calculations were performed on SOM2 and SOM3 at IFIC 
and on the COSMOS Consortium supercomputer within the DiRAC 
Facility jointly funded by STFC, the Large Facilities Capital Fund of BIS and the
University of Cambridge, as well as the Darwin Supercomputer of the
University of Cambridge High Performance Computing Service (http://
www.hpc.cam.ac.uk/), provided by Dell Inc. using Strategic Research
Infrastructure Funding from the Higher Education Funding Council for
England.  FVN and MV are supported by the ERC Starting Grant
``cosmoIGM''. MV is also supported by I.S. INFN/PD51. 
ES was supported in part by NSF-AST 0908241. 
SS is supported by a Grant-in-Aid for Young Scientists (Start-up) 
from the Japan Society for the Promotion of Science (JSPS) (No. 25887012).

\appendix

\section{Power spectrum corrections and errors}
\label{Appendix_A}

The power spectrum of a distribution of points that arise from a
Poisson sampling of a underlying density field is related with that of
the underlying field by (see for instance \cite{Smith_2009})

\begin{equation}
P^d(\vec{k})=P^c(\vec{k})+\frac{1}{\bar{n}}~,
\label{shot-noise}
\end{equation}
where $\bar{n}$ is the points number density and $P^d(\vec{k})$ and
$P^c(\vec{k})$ are the power spectra of the discrete point set and of
the continuos field, respectively. However, when computing the power
spectrum of the CDM particles we should not subtract the
$\bar{n}^{-1}$ term (we will refer to this extra contribution in the power
spectrum as shot-noise). This is because we generated the initial
conditions of the CDM particles on a grid, perturbing their positions
around the grid points according to the gravitational potential, and
this process does not correspond to a Poisson sampling of the density
field.

Although the initial conditions of the neutrino particles were also
generated on a grid, soon after the simulation starts, the large
neutrino thermal velocities make them to cross the simulation box
several times, forcing neutrinos to effectively distribute randomly
over the simulation box with a value of their power spectrum 
equal to $\bar{n}^{-1}$ on all
scales. Therefore, when computing the neutrino power spectrum we
correct for the neutrino shot-noise by using eq. \ref{shot-noise}. The
halo distribution is expected to show up as a Poisson process over the
underlying matter density field. Thus, we have corrected the halo
power spectrum to account for its shot-noise.

When there is more than one fluid, for instance CDM and neutrinos, the
total matter power spectrum is related to the individual ones and to
their cross-power spectrum by

\begin{equation}
P_{\rm m} (\vec{k})= \left(\frac{\Omega_{\rm cdm}}{\Omega_{\rm 
m}}\right)^2P_{\rm 
cdm}(\vec{k})+\left(\frac{\Omega_\nu}{\Omega_{\rm 
m}}\right)^2P_\nu(\vec{k})+\left(\frac{2~\Omega_{\rm cdm}~\Omega_\nu}{\Omega_{\rm m}^2}\right)P_{\rm 
cdm-\nu}(\vec{k})~,
\end{equation}
where $\Omega_{\rm m}=\Omega_{\rm cdm}+\Omega_\nu$. As we have seen,
the CDM power spectrum does not suffer from shot-noise, and moreover,
since the positions of the neutrino and CDM particles do not overlap,
the CDM-neutrinos cross-power spectrum is not affected by shot-noise
\cite{Smith_2009}. Therefore, the shot-noise of the total matter power spectrum
is given by $\left(\frac{\Omega_\nu}{\Omega_{\rm
    m}}\right)^2\bar{n}_\nu^{-1}$. We note that this quantity is very
small, and thus, the correction is, on practice, negligible. Therefore, for simplicity,
we decide to ignore the neutrino shot-noise contribution to
 the total matter power spectrum.

The halo-matter and halo-cold dark matter cross-power spectrums are
 affected by shot-noise because
the halo positions are a subsample of the positions of the CDM
particles\footnote{This is because our halo finder assigns the halo
  center to the position of particle in which the gravitational
  potential is minimum.}. The shot-noise level of the cross-power
spectrum is given in this case by shot-noise of the matter
distribution (see \cite{Smith_2009} for details), which, as we have
seen, is negligible and thus, we do not correct for it.

Finally, since the simulations we have used to compute the bias in Fourier space
contains 8 independent realizations, we compute the errors on the measurements 
of both the power spectrum and 
the bias just by calculating the standard deviation around the mean value from the whole set.

\section{Impact of the halo identification criteria on the halo-matter bias}
\label{Appendix_C}

Here we investigate whether our results, in terms of the bias between the spatial distribution of
dark matter haloes and that of the underlying matter, are affected by our simplified assumption 
of running the FoF and SUBFIND algorithms just on top of the CDM distribution. In order to 
test the robustness of our results, we run the SUBFIND algorithm over the total matter 
distribution, i.e. on top of the CDM 
plus the neutrino distribution, of the simulations H0, H3 and H6 to identify the SO haloes. 
We then compute the bias, $b_{\rm hm}$ and 
$b_{\rm hh}$, for each of the 8 independent realizations for
each cosmological model at redshifts 
$z=0$, $z=0.5$ and $z=1$ using haloes with masses, $M_{200}$, above 
$2\times10^{13}$ $h^{-1}$M$_\odot$ and $4\times10^{13}$ $h^{-1}$M$_\odot$. 
Finally, for each cosmological model, we calculate the mean bias from the results of the 8 
independent realizations.

\begin{figure}
\begin{center}
\includegraphics[width=1.0\textwidth]{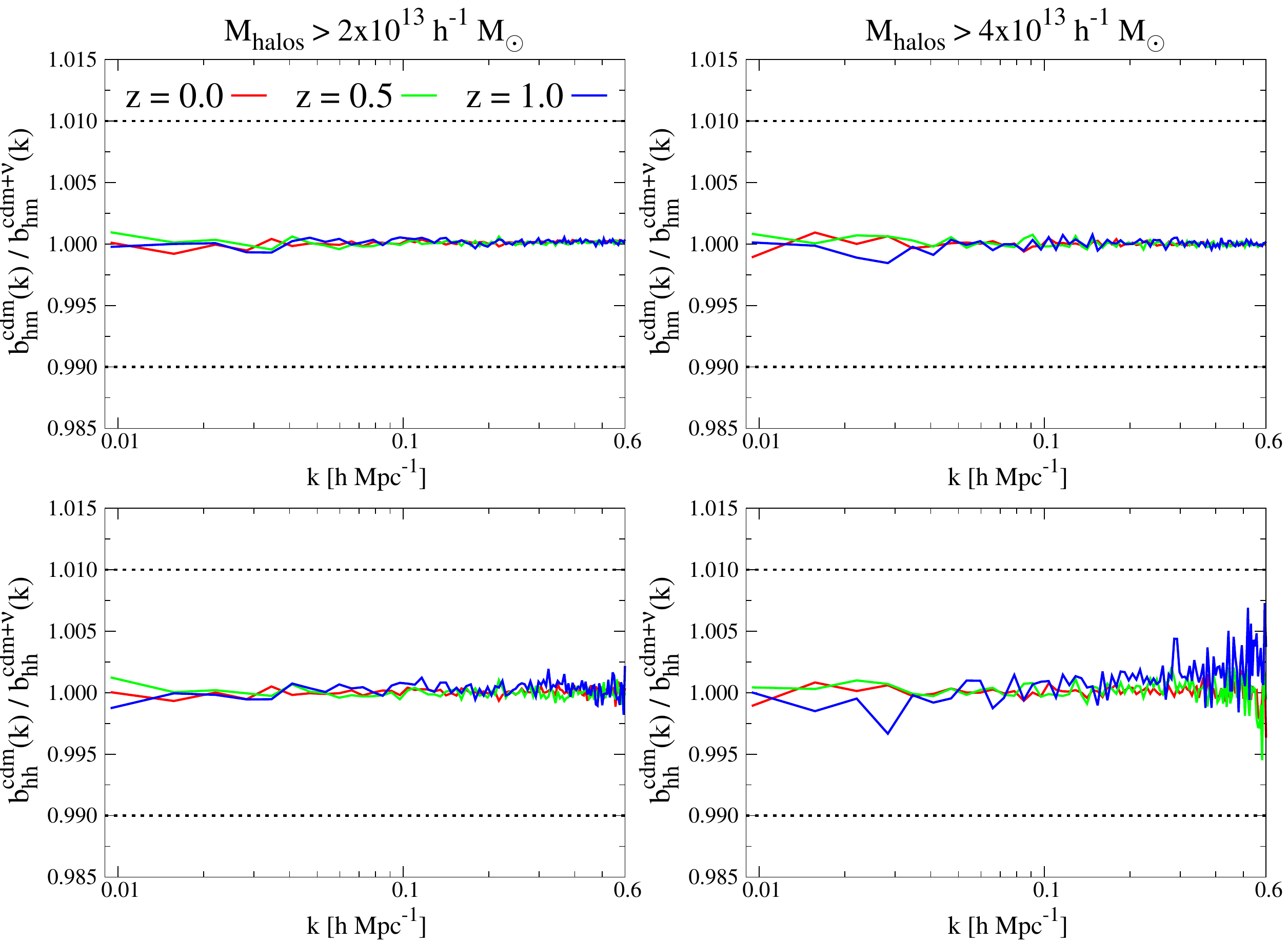}
\end{center}
\caption{Robustness of our bias results with respect to the halo identification procedure. We 
compute the bias, $b_{\rm hm}(k)$ and $b_{\rm hh}(k)$ by selecting the SO haloes over the
total matter distribution. In the panels we plot the ratio of the bias calculated using SO haloes
identified over the CDM distribution alone over the bias computed by selecting SO haloes
on top of the total matter distribution. The left column shows the results for the haloes with masses,
$M_{200}$, above $2\times10^{13}$ $h^{-1}$M$_\odot$ whereas the right column is for haloes
with masses above $4\times10^{13}$ $h^{-1}$M$_\odot$. The upper panels show the results for
the bias computed as $b_{\rm hm}(k)$ whereas the bottom panels are for the bias computed as
$b_{\rm hh}(k)$. Results are displayed at $z=0$ (red), $z=0.5$ (green) and $z=1$ (blue). The 
horizontal dotted lines represent deviations of $1\%$.}
\label{error_Pk}
\end{figure}

In Fig. \ref{error_Pk} we show the ratio of the bias computed by using the SO haloes found 
on top of the CDM distribution alone to the bias computed by using SO haloes identified on 
top of the total matter distribution. The left column shows the results for haloes with 
masses above  $2\times10^{13}$ $h^{-1}$M$_\odot$ whereas the right column displays 
the results for haloes
with masses larger than $4\times10^{13}$ $h^{-1}$M$_\odot$. The upper panels show the 
ratio using the bias computed as $b_{\rm hm}(k)$ while the bottom panels are for bias computed
as $b_{\rm hh}(k)$. We find that by selecting the dark matter haloes on top of the total matter
distribution, our results, in terms of the bias, are basically unchanged. The difference between 
computing the bias with dark matter haloes selected from the CDM distribution alone and computing
the bias with haloes identified from the total matter distribution is less than $\sim0.5\%$ for the 
values of $k$ considered in this work. We conclude that our results are not affected by
our simplified assumption of using dark matter haloes as identified from the CDM distribution alone.

\bibliographystyle{JHEP}
\bibliography{Bibliography}

\providecommand{\href}[2]{#2}\begingroup\raggedright\begin{thebibliography}{10}

\bibitem{CowanReines}
C.~L. {Cowan}, Jr., F.~{Reines}, F.~B. {Harrison}, H.~W. {Kruse}, and A.~D.
  {McGuire}, {\it {Detection of the Free Neutrino: A Confirmation}},  {\em
  Science} {\bf 124} (July, 1956) 103--104.

\bibitem{Cleveland}
B.~T. {Cleveland}, T.~{Daily}, R.~{Davis}, Jr., J.~R. {Distel}, K.~{Lande},
  C.~K. {Lee}, P.~S. {Wildenhain}, and J.~{Ullman}, {\it {Measurement of the
  Solar Electron Neutrino Flux with the Homestake Chlorine Detector}},  {\em
  \apj} {\bf 496} (Mar., 1998) 505.

\bibitem{Fogli}
G.~L. {Fogli}, E.~{Lisi}, A.~{Marrone}, D.~{Montanino}, A.~{Palazzo}, and A.~M.
  {Rotunno}, {\it {Global analysis of neutrino masses, mixings, and phases:
  Entering the era of leptonic CP violation searches}},  {\em \prd} {\bf 86}
  (July, 2012) 013012, [\href{http://xxx.lanl.gov/abs/1205.5254}{{\tt
  arXiv:1205.5254}}].

\bibitem{Tortola}
D.~V. {Forero}, M.~{T{\'o}rtola}, and J.~W.~F. {Valle}, {\it {Global status of
  neutrino oscillation parameters after Neutrino-2012}},  {\em \prd} {\bf 86}
  (Oct., 2012) 073012, [\href{http://xxx.lanl.gov/abs/1205.4018}{{\tt
  arXiv:1205.4018}}].

\bibitem{LesgourguesPastor}
J.~{Lesgourgues} and S.~{Pastor}, {\it {Massive neutrinos and cosmology}},
  {\em \physrep} {\bf 429} (July, 2006) 307--379,
  [\href{http://xxx.lanl.gov/abs/astro-ph/}{{\tt astro-ph/}}].

\bibitem{Hannestad_2003}
S.~{Hannestad}, {\it {Neutrino masses and the number of neutrino species from
  WMAP and 2dFGRS}},  {\em \jcap} {\bf 5} (May, 2003) 4,
  [\href{http://xxx.lanl.gov/abs/astro-ph/}{{\tt astro-ph/}}].

\bibitem{Reid}
B.~A. {Reid}, L.~{Verde}, R.~{Jimenez}, and O.~{Mena}, {\it {Robust neutrino
  constraints by combining low redshift observations with the CMB}},  {\em
  \jcap} {\bf 1} (Jan., 2010) 3, [\href{http://xxx.lanl.gov/abs/0910.0008}{{\tt
  arXiv:0910.0008}}].

\bibitem{Thomas}
S.~A. {Thomas}, F.~B. {Abdalla}, and O.~{Lahav}, {\it {Upper Bound of 0.28 eV
  on Neutrino Masses from the Largest Photometric Redshift Survey}},  {\em
  Physical Review Letters} {\bf 105} (July, 2010) 031301,
  [\href{http://xxx.lanl.gov/abs/0911.5291}{{\tt arXiv:0911.5291}}].

\bibitem{Swanson}
M.~E.~C. {Swanson}, W.~J. {Percival}, and O.~{Lahav}, {\it {Neutrino masses
  from clustering of red and blue galaxies: a test of astrophysical
  uncertainties}},  {\em \mnras} {\bf 409} (Dec., 2010) 1100--1112,
  [\href{http://xxx.lanl.gov/abs/1006.2825}{{\tt arXiv:1006.2825}}].

\bibitem{Saito_2010}
S.~{Saito}, M.~{Takada}, and A.~{Taruya}, {\it {Neutrino mass constraint from
  the Sloan Digital Sky Survey power spectrum of luminous red galaxies and
  perturbation theory}},  {\em \prd} {\bf 83} (Feb., 2011) 043529,
  [\href{http://xxx.lanl.gov/abs/1006.4845}{{\tt arXiv:1006.4845}}].

\bibitem{dePutter}
R.~{de Putter}, O.~{Mena}, E.~{Giusarma}, S.~{Ho}, A.~{Cuesta}, H.-J. {Seo},
  A.~J. {Ross}, M.~{White}, D.~{Bizyaev}, H.~{Brewington}, D.~{Kirkby},
  E.~{Malanushenko}, V.~{Malanushenko}, D.~{Oravetz}, K.~{Pan}, W.~J.
  {Percival}, N.~P. {Ross}, D.~P. {Schneider}, A.~{Shelden}, A.~{Simmons}, and
  S.~{Snedden}, {\it {New Neutrino Mass Bounds from SDSS-III Data Release 8
  Photometric Luminous Galaxies}},  {\em \apj} {\bf 761} (Dec., 2012) 12,
  [\href{http://xxx.lanl.gov/abs/1201.1909}{{\tt arXiv:1201.1909}}].

\bibitem{Xia2012}
J.-Q. {Xia}, B.~R. {Granett}, M.~{Viel}, S.~{Bird}, L.~{Guzzo}, M.~G.
  {Haehnelt}, J.~{Coupon}, H.~J. {McCracken}, and Y.~{Mellier}, {\it
  {Constraints on massive neutrinos from the CFHTLS angular power spectrum}},
  {\em \jcap} {\bf 6} (June, 2012) 10,
  [\href{http://xxx.lanl.gov/abs/1203.5105}{{\tt arXiv:1203.5105}}].

\bibitem{WiggleZ}
S.~{Riemer-S{\o}rensen}, C.~{Blake}, D.~{Parkinson}, T.~M. {Davis},
  S.~{Brough}, M.~{Colless}, C.~{Contreras}, W.~{Couch}, S.~{Croom},
  D.~{Croton}, M.~J. {Drinkwater}, K.~{Forster}, D.~{Gilbank}, M.~{Gladders},
  K.~{Glazebrook}, B.~{Jelliffe}, R.~J. {Jurek}, I.-h. {Li}, B.~{Madore}, D.~C.
  {Martin}, K.~{Pimbblet}, G.~B. {Poole}, M.~{Pracy}, R.~{Sharp},
  E.~{Wisnioski}, D.~{Woods}, T.~K. {Wyder}, and H.~K.~C. {Yee}, {\it {WiggleZ
  Dark Energy Survey: Cosmological neutrino mass constraint from blue
  high-redshift galaxies}},  {\em \prd} {\bf 85} (Apr., 2012) 081101,
  [\href{http://xxx.lanl.gov/abs/1112.4940}{{\tt arXiv:1112.4940}}].

\bibitem{Zhao2012}
G.-B. {Zhao}, S.~{Saito}, W.~J. {Percival}, A.~J. {Ross}, F.~{Montesano},
  M.~{Viel}, D.~P. {Schneider}, D.~J. {Ernst}, M.~{Manera},
  J.~{Miralda-Escude}, N.~P. {Ross}, L.~{Samushia}, A.~G. {Sanchez}, M.~E.~C.
  {Swanson}, D.~{Thomas}, R.~{Tojeiro}, C.~{Yeche}, and D.~G. {York}, {\it {The
  clustering of galaxies in the SDSS-III Baryon Oscillation Spectroscopic
  Survey: weighing the neutrino mass using the galaxy power spectrum of the
  CMASS sample}},  {\em ArXiv e-prints} (Nov., 2012)
  [\href{http://xxx.lanl.gov/abs/1211.3741}{{\tt arXiv:1211.3741}}].

\bibitem{2013MNRAS.430..747M}
S.~{More}, F.~C. {van den Bosch}, M.~{Cacciato}, A.~{More}, H.~{Mo}, and
  X.~{Yang}, {\it {Cosmological constraints from a combination of galaxy
  clustering and lensing - II. Fisher matrix analysis}},  {\em \mnras} {\bf
  430} (Apr., 2013) 747--766, [\href{http://xxx.lanl.gov/abs/1207.0004}{{\tt
  arXiv:1207.0004}}].

\bibitem{Riemer-Sorensen}
S.~{Riemer-S{\o}rensen}, D.~{Parkinson}, and T.~M. {Davis}, {\it {Combining
  Planck with Large Scale Structure gives strong neutrino mass constraint}},
  {\em ArXiv e-prints} (June, 2013)
  [\href{http://xxx.lanl.gov/abs/1306.4153}{{\tt arXiv:1306.4153}}].

\bibitem{Costanzi}
M.~{Costanzi Alunno Cerbolini}, B.~{Sartoris}, J.-Q. {Xia}, A.~{Biviano},
  S.~{Borgani}, and M.~{Viel}, {\it {Constraining neutrino properties with a
  Euclid-like galaxy cluster survey}},  {\em \jcap} {\bf 6} (June, 2013) 20,
  [\href{http://xxx.lanl.gov/abs/1303.4550}{{\tt arXiv:1303.4550}}].

\bibitem{Basse}
T.~{Basse}, O.~{Eggers Bjaelde}, J.~{Hamann}, S.~{Hannestad}, and Y.~Y.~Y.
  {Wong}, {\it {Dark energy and neutrino constraints from a future EUCLID-like
  survey}},  {\em ArXiv e-prints} (Apr., 2013)
  [\href{http://xxx.lanl.gov/abs/1304.2321}{{\tt arXiv:1304.2321}}].

\bibitem{carbone2011}
C.~{Carbone}, L.~{Verde}, Y.~{Wang}, and A.~{Cimatti}, {\it {Neutrino
  constraints from future nearly all-sky spectroscopic galaxy surveys}},  {\em
  \jcap} {\bf 3} (Mar., 2011) 30,
  [\href{http://xxx.lanl.gov/abs/1012.2868}{{\tt arXiv:1012.2868}}].

\bibitem{Planck_2013}
{Planck Collaboration}, P.~A.~R. {Ade}, N.~{Aghanim}, C.~{Armitage-Caplan},
  M.~{Arnaud}, M.~{Ashdown}, F.~{Atrio-Barandela}, J.~{Aumont},
  C.~{Baccigalupi}, A.~J. {Banday}, and et~al., {\it {Planck 2013 results. XVI.
  Cosmological parameters}},  {\em ArXiv e-prints} (Mar., 2013)
  [\href{http://xxx.lanl.gov/abs/1303.5076}{{\tt arXiv:1303.5076}}].

\bibitem{LesgourguesBook}
J.~{Lesgourgues}, G.~{Mangano}, G.~{Miele}, and {Pastor}, {\em {Neutrino
  cosmology}}.
\newblock Apr., 2013.

\bibitem{Ma}
S.~{Singh} and C.-P. {Ma}, {\it {Neutrino clustering in cold dark matter halos:
  Implications for ultrahigh energy cosmic rays}},  {\em \prd} {\bf 67} (Jan.,
  2003) 023506, [\href{http://xxx.lanl.gov/abs/astro-ph/}{{\tt astro-ph/}}].

\bibitem{Wong}
A.~{Ringwald} and Y.~Y.~Y. {Wong}, {\it {Gravitational clustering of relic
  neutrinos and implications for their detection}},  {\em \jcap} {\bf 12}
  (Dec., 2004) 5, [\href{http://xxx.lanl.gov/abs/hep-ph/04}{{\tt hep-ph/04}}].

\bibitem{Brandbyge_haloes}
J.~{Brandbyge}, S.~{Hannestad}, T.~{Haugb{\o}lle}, and Y.~Y.~Y. {Wong}, {\it
  {Neutrinos in non-linear structure formation - the effect on halo
  properties}},  {\em \jcap} {\bf 9} (Sept., 2010) 14,
  [\href{http://xxx.lanl.gov/abs/1004.4105}{{\tt arXiv:1004.4105}}].

\bibitem{Villaescusa-Navarro_2011}
F.~{Villaescusa-Navarro}, J.~{Miralda-Escud{\'e}}, C.~{Pe{\~n}a-Garay}, and
  V.~{Quilis}, {\it {Neutrino halos in clusters of galaxies and their weak
  lensing signature}},  {\em \jcap} {\bf 6} (June, 2011) 27,
  [\href{http://xxx.lanl.gov/abs/1104.4770}{{\tt arXiv:1104.4770}}].

\bibitem{Villaescusa-Navarro_2013a}
F.~{Villaescusa-Navarro}, S.~{Bird}, C.~{Pe{\~n}a-Garay}, and M.~{Viel}, {\it
  {Non-linear evolution of the cosmic neutrino background}},  {\em \jcap} {\bf
  3} (Mar., 2013) 19, [\href{http://xxx.lanl.gov/abs/1212.4855}{{\tt
  arXiv:1212.4855}}].

\bibitem{Brandbyge2008}
J.~{Brandbyge}, S.~{Hannestad}, T.~{Haugb{\o}lle}, and B.~{Thomsen}, {\it {The
  effect of thermal neutrino motion on the non-linear cosmological matter power
  spectrum}},  {\em \jcap} {\bf 8} (Aug., 2008) 20,
  [\href{http://xxx.lanl.gov/abs/0802.3700}{{\tt arXiv:0802.3700}}].

\bibitem{Saito_2008}
S.~{Saito}, M.~{Takada}, and A.~{Taruya}, {\it {Impact of Massive Neutrinos on
  the Nonlinear Matter Power Spectrum}},  {\em Physical Review Letters} {\bf
  100} (May, 2008) 191301, [\href{http://xxx.lanl.gov/abs/0801.0607}{{\tt
  arXiv:0801.0607}}].

\bibitem{Brandbyge2009}
J.~{Brandbyge} and S.~{Hannestad}, {\it {Grid based linear neutrino
  perturbations in cosmological N-body simulations}},  {\em \jcap} {\bf 5}
  (May, 2009) 2, [\href{http://xxx.lanl.gov/abs/0812.3149}{{\tt
  arXiv:0812.3149}}].

\bibitem{BrandbygeHybrid}
J.~{Brandbyge} and S.~{Hannestad}, {\it {Resolving cosmic neutrino structure: a
  hybrid neutrino N-body scheme}},  {\em \jcap} {\bf 1} (Jan., 2010) 21,
  [\href{http://xxx.lanl.gov/abs/0908.1969}{{\tt arXiv:0908.1969}}].

\bibitem{Saito_2009}
S.~{Saito}, M.~{Takada}, and A.~{Taruya}, {\it {Nonlinear power spectrum in the
  presence of massive neutrinos: Perturbation theory approach, galaxy bias, and
  parameter forecasts}},  {\em \prd} {\bf 80} (Oct., 2009) 083528,
  [\href{http://xxx.lanl.gov/abs/0907.2922}{{\tt arXiv:0907.2922}}].

\bibitem{Viel_2010}
M.~{Viel}, M.~G. {Haehnelt}, and V.~{Springel}, {\it {The effect of neutrinos
  on the matter distribution as probed by the intergalactic medium}},  {\em
  \jcap} {\bf 6} (June, 2010) 15,
  [\href{http://xxx.lanl.gov/abs/1003.2422}{{\tt arXiv:1003.2422}}].

\bibitem{Agarwal2011}
S.~{Agarwal} and H.~A. {Feldman}, {\it {The effect of massive neutrinos on the
  matter power spectrum}},  {\em \mnras} {\bf 410} (Jan., 2011) 1647--1654,
  [\href{http://xxx.lanl.gov/abs/1006.0689}{{\tt arXiv:1006.0689}}].

\bibitem{Bird_2011}
S.~{Bird}, M.~{Viel}, and M.~G. {Haehnelt}, {\it {Massive neutrinos and the
  non-linear matter power spectrum}},  {\em \mnras} {\bf 420} (Mar., 2012)
  2551--2561, [\href{http://xxx.lanl.gov/abs/1109.4416}{{\tt
  arXiv:1109.4416}}].

\bibitem{Wagner2012}
C.~{Wagner}, L.~{Verde}, and R.~{Jimenez}, {\it {Effects of the Neutrino Mass
  Splitting on the Nonlinear Matter Power Spectrum}},  {\em \apjl} {\bf 752}
  (June, 2012) L31, [\href{http://xxx.lanl.gov/abs/1203.5342}{{\tt
  arXiv:1203.5342}}].

\bibitem{Villaescusa-Navarro_2012}
F.~{Villaescusa-Navarro}, M.~{Vogelsberger}, M.~{Viel}, and A.~{Loeb}, {\it
  {Neutrino Signatures on the High Transmission Regions of the Lyman-alpha
  Forest}},  {\em ArXiv e-prints} (June, 2011)
  [\href{http://xxx.lanl.gov/abs/1106.2543}{{\tt arXiv:1106.2543}}].

\bibitem{Marulli_2011}
F.~{Marulli}, C.~{Carbone}, M.~{Viel}, L.~{Moscardini}, and A.~{Cimatti}, {\it
  {Effects of massive neutrinos on the large-scale structure of the Universe}},
   {\em \mnras} {\bf 418} (Nov., 2011) 346--356,
  [\href{http://xxx.lanl.gov/abs/1103.0278}{{\tt arXiv:1103.0278}}].

\bibitem{2013arXiv1310.7571V}
M.~P. {van Daalen}, J.~{Schaye}, I.~G. {McCarthy}, C.~M. {Booth}, and C.~{Dalla
  Vecchia}, {\it {The impact of baryonic processes on the two-point correlation
  functions of galaxies, subhaloes and matter}},  {\em ArXiv e-prints} (Oct.,
  2013) [\href{http://xxx.lanl.gov/abs/1310.7571}{{\tt arXiv:1310.7571}}].

\bibitem{Castorina}
E.~{Castorina} and et~al., {\it {Neutrino mass scale-dependence bias}},  {\em
  \jcap} (Oct., 2013) X, [\href{http://xxx.lanl.gov/abs/1310.XXXX}{{\tt
  arXiv:1310.XXXX}}].

\bibitem{Costanzi2}
M.~{Costanzi} and et~al., {\it {The halo mass function in a universe with
  massive neutrinos: application to galaxy clusters}},  {\em \jcap} (Oct.,
  2013) X, [\href{http://xxx.lanl.gov/abs/1310.XXXX}{{\tt arXiv:1310.XXXX}}].

\bibitem{Crocce_2010}
M.~{Crocce}, P.~{Fosalba}, F.~J. {Castander}, and E.~{Gazta{\~n}aga}, {\it
  {Simulating the Universe with MICE: the abundance of massive clusters}},
  {\em \mnras} {\bf 403} (Apr., 2010) 1353--1367,
  [\href{http://xxx.lanl.gov/abs/0907.0019}{{\tt arXiv:0907.0019}}].

\bibitem{Tinker_2008}
J.~{Tinker}, A.~V. {Kravtsov}, A.~{Klypin}, K.~{Abazajian}, M.~{Warren},
  G.~{Yepes}, S.~{Gottl{\"o}ber}, and D.~E. {Holz}, {\it {Toward a Halo Mass
  Function for Precision Cosmology: The Limits of Universality}},  {\em \apj}
  {\bf 688} (Dec., 2008) 709--728,
  [\href{http://xxx.lanl.gov/abs/0803.2706}{{\tt arXiv:0803.2706}}].

\bibitem{2012arXiv1212.6267H}
Z.~{Hou}, C.~L. {Reichardt}, K.~T. {Story}, B.~{Follin}, R.~{Keisler}, K.~A.
  {Aird}, B.~A. {Benson}, L.~E. {Bleem}, J.~E. {Carlstrom}, C.~L. {Chang},
  H.~{Cho}, T.~M. {Crawford}, A.~T. {Crites}, T.~{de Haan}, R.~{de Putter},
  M.~A. {Dobbs}, and S.~{Dodelson}, {\it {Constraints on Cosmology from the
  Cosmic Microwave Background Power Spectrum of the 2500-square degree SPT-SZ
  Survey}},  {\em ArXiv e-prints} (Dec., 2012)
  [\href{http://xxx.lanl.gov/abs/1212.6267}{{\tt arXiv:1212.6267}}].

\bibitem{Springel_2005}
V.~{Springel}, {\it {The cosmological simulation code GADGET-2}},  {\em \mnras}
  {\bf 364} (Dec., 2005) 1105--1134,
  [\href{http://xxx.lanl.gov/abs/astro-ph/}{{\tt astro-ph/}}].

\bibitem{CAMB}
A.~{Lewis}, A.~{Challinor}, and A.~{Lasenby}, {\it {Efficient Computation of
  Cosmic Microwave Background Anisotropies in Closed Friedmann-Robertson-Walker
  Models}},  {\em \apj} {\bf 538} (Aug., 2000) 473--476,
  [\href{http://xxx.lanl.gov/abs/astro-ph/}{{\tt astro-ph/}}].

\bibitem{FoF}
M.~{Davis}, G.~{Efstathiou}, C.~S. {Frenk}, and S.~D.~M. {White}, {\it {The
  evolution of large-scale structure in a universe dominated by cold dark
  matter}},  {\em \apj} {\bf 292} (May, 1985) 371--394.

\bibitem{Subfind}
V.~{Springel}, S.~D.~M. {White}, G.~{Tormen}, and G.~{Kauffmann}, {\it
  {Populating a cluster of galaxies - I. Results at [formmu2]z=0}},  {\em
  \mnras} {\bf 328} (Dec., 2001) 726--750,
  [\href{http://xxx.lanl.gov/abs/astro-ph/}{{\tt astro-ph/}}].

\bibitem{Dekel}
A.~{Dekel} and O.~{Lahav}, {\it {Stochastic Nonlinear Galaxy Biasing}},  {\em
  \apj} {\bf 520} (July, 1999) 24--34,
  [\href{http://xxx.lanl.gov/abs/astro-ph/}{{\tt astro-ph/}}].

\bibitem{Hamaus_2010}
N.~{Hamaus}, U.~{Seljak}, V.~{Desjacques}, R.~E. {Smith}, and T.~{Baldauf},
  {\it {Minimizing the stochasticity of halos in large-scale structure
  surveys}},  {\em \prd} {\bf 82} (Aug., 2010) 043515,
  [\href{http://xxx.lanl.gov/abs/1004.5377}{{\tt arXiv:1004.5377}}].

\bibitem{Baldauf_2013}
T.~{Baldauf}, U.~{Seljak}, R.~E. {Smith}, N.~{Hamaus}, and V.~{Desjacques},
  {\it {Halo Stochasticity from Exclusion and non-linear Clustering}},  {\em
  ArXiv e-prints} (May, 2013) [\href{http://xxx.lanl.gov/abs/1305.2917}{{\tt
  arXiv:1305.2917}}].

\bibitem{Smith_2006}
R.~E. {Smith}, R.~{Scoccimarro}, and R.~K. {Sheth}, {\it {Scale dependence of
  halo and galaxy bias: Effects in real space}},  {\em \prd} {\bf 75} (Mar.,
  2007) 063512, [\href{http://xxx.lanl.gov/abs/astro-ph/}{{\tt astro-ph/}}].

\bibitem{Baldauf_2009}
T.~{Baldauf}, R.~E. {Smith}, U.~{Seljak}, and R.~{Mandelbaum}, {\it {Algorithm
  for the direct reconstruction of the dark matter correlation function from
  weak lensing and galaxy clustering}},  {\em \prd} {\bf 81} (Mar., 2010)
  063531, [\href{http://xxx.lanl.gov/abs/0911.4973}{{\tt arXiv:0911.4973}}].

\bibitem{Jing_2005}
Y.~P. {Jing}, {\it {Correcting for the Alias Effect When Measuring the Power
  Spectrum Using a Fast Fourier Transform}},  {\em \apj} {\bf 620} (Feb., 2005)
  559--563, [\href{http://xxx.lanl.gov/abs/astro-ph/}{{\tt astro-ph/}}].

\bibitem{Montesano_2010}
F.~{Montesano}, A.~G. {S{\'a}nchez}, and S.~{Phleps}, {\it {A new model for the
  full shape of the large-scale power spectrum}},  {\em \mnras} {\bf 408}
  (Nov., 2010) 2397--2412, [\href{http://xxx.lanl.gov/abs/1007.0755}{{\tt
  arXiv:1007.0755}}].

\bibitem{Tinker_2010}
J.~L. {Tinker}, B.~E. {Robertson}, A.~V. {Kravtsov}, A.~{Klypin}, M.~S.
  {Warren}, G.~{Yepes}, and S.~{Gottl{\"o}ber}, {\it {The Large-scale Bias of
  Dark Matter Halos: Numerical Calibration and Model Tests}},  {\em \apj} {\bf
  724} (Dec., 2010) 878--886, [\href{http://xxx.lanl.gov/abs/1001.3162}{{\tt
  arXiv:1001.3162}}].

\bibitem{Ichiki-Takada}
K.~{Ichiki} and M.~{Takada}, {\it {Impact of massive neutrinos on the abundance
  of massive clusters}},  {\em \prd} {\bf 85} (Mar., 2012) 063521,
  [\href{http://xxx.lanl.gov/abs/1108.4688}{{\tt arXiv:1108.4688}}].

\bibitem{Landy-Szalay_93}
S.~D. {Landy} and A.~S. {Szalay}, {\it {Bias and variance of angular
  correlation functions}},  {\em \apj} {\bf 412} (July, 1993) 64--71.

\bibitem{Szapudi}
I.~{Szapudi} and A.~S. {Szalay}, {\it {A New Class of Estimators for the
  N-point Correlations}},  {\em ArXiv Astrophysics e-prints} (Apr., 1997)
  [\href{http://xxx.lanl.gov/abs/astro-ph/}{{\tt astro-ph/}}].

\bibitem{Ma_Fry_2000}
C.-P. {Ma} and J.~N. {Fry}, {\it {Deriving the Nonlinear Cosmological Power
  Spectrum and Bispectrum from Analytic Dark Matter Halo Profiles and Mass
  Functions}},  {\em \apj} {\bf 543} (Nov., 2000) 503--513,
  [\href{http://xxx.lanl.gov/abs/astro-ph/}{{\tt astro-ph/}}].

\bibitem{Peacock_Smith_2000}
J.~A. {Peacock} and R.~E. {Smith}, {\it {Halo occupation numbers and galaxy
  bias}},  {\em \mnras} {\bf 318} (Nov., 2000) 1144--1156,
  [\href{http://xxx.lanl.gov/abs/astro-ph/}{{\tt astro-ph/}}].

\bibitem{Seljak_2000}
U.~{Seljak}, {\it {Analytic model for galaxy and dark matter clustering}},
  {\em \mnras} {\bf 318} (Oct., 2000) 203--213,
  [\href{http://xxx.lanl.gov/abs/astro-ph/}{{\tt astro-ph/}}].

\bibitem{Scoccimarro_2001}
R.~{Scoccimarro}, R.~K. {Sheth}, L.~{Hui}, and B.~{Jain}, {\it {How Many
  Galaxies Fit in a Halo? Constraints on Galaxy Formation Efficiency from
  Spatial Clustering}},  {\em \apj} {\bf 546} (Jan., 2001) 20--34,
  [\href{http://xxx.lanl.gov/abs/astro-ph/}{{\tt astro-ph/}}].

\bibitem{Berlind_2002}
A.~A. {Berlind} and D.~H. {Weinberg}, {\it {The Halo Occupation Distribution:
  Toward an Empirical Determination of the Relation between Galaxies and
  Mass}},  {\em \apj} {\bf 575} (Aug., 2002) 587--616,
  [\href{http://xxx.lanl.gov/abs/astro-ph/}{{\tt astro-ph/}}].

\bibitem{Kravtsov_2004}
A.~V. {Kravtsov}, A.~A. {Berlind}, R.~H. {Wechsler}, A.~A. {Klypin},
  S.~{Gottl{\"o}ber}, B.~{Allgood}, and J.~R. {Primack}, {\it {The Dark Side of
  the Halo Occupation Distribution}},  {\em \apj} {\bf 609} (July, 2004)
  35--49, [\href{http://xxx.lanl.gov/abs/astro-ph/}{{\tt astro-ph/}}].

\bibitem{Zheng_2004}
Z.~{Zheng}, A.~A. {Berlind}, D.~H. {Weinberg}, A.~J. {Benson}, C.~M. {Baugh},
  S.~{Cole}, R.~{Dav{\'e}}, C.~S. {Frenk}, N.~{Katz}, and C.~G. {Lacey}, {\it
  {Theoretical Models of the Halo Occupation Distribution: Separating Central
  and Satellite Galaxies}},  {\em \apj} {\bf 633} (Nov., 2005) 791--809,
  [\href{http://xxx.lanl.gov/abs/astro-ph/}{{\tt astro-ph/}}].

\bibitem{Zehavi_2011}
I.~{Zehavi}, Z.~{Zheng}, D.~H. {Weinberg}, M.~R. {Blanton}, N.~A. {Bahcall},
  A.~A. {Berlind}, J.~{Brinkmann}, J.~A. {Frieman}, J.~E. {Gunn}, R.~H.
  {Lupton}, R.~C. {Nichol}, W.~J. {Percival}, D.~P. {Schneider}, R.~A.
  {Skibba}, M.~A. {Strauss}, M.~{Tegmark}, and D.~G. {York}, {\it {Galaxy
  Clustering in the Completed SDSS Redshift Survey: The Dependence on Color and
  Luminosity}},  {\em \apj} {\bf 736} (July, 2011) 59,
  [\href{http://xxx.lanl.gov/abs/1005.2413}{{\tt arXiv:1005.2413}}].

\bibitem{Blanton_2003}
M.~R. {Blanton}, D.~W. {Hogg}, N.~A. {Bahcall}, J.~{Brinkmann}, M.~{Britton},
  A.~J. {Connolly}, I.~{Csabai}, M.~{Fukugita}, J.~{Loveday}, A.~{Meiksin},
  J.~A. {Munn}, R.~C. {Nichol}, S.~{Okamura}, T.~{Quinn}, D.~P. {Schneider},
  K.~{Shimasaku}, M.~A. {Strauss}, M.~{Tegmark}, M.~S. {Vogeley}, and D.~H.
  {Weinberg}, {\it {The Galaxy Luminosity Function and Luminosity Density at
  Redshift z = 0.1}},  {\em \apj} {\bf 592} (Aug., 2003) 819--838,
  [\href{http://xxx.lanl.gov/abs/astro-ph/}{{\tt astro-ph/}}].

\bibitem{Zehavi_2004}
I.~{Zehavi}, Z.~{Zheng}, D.~H. {Weinberg}, J.~A. {Frieman}, A.~A. {Berlind},
  M.~R. {Blanton}, R.~{Scoccimarro}, R.~K. {Sheth}, M.~A. {Strauss}, I.~{Kayo},
  Y.~{Suto}, M.~{Fukugita}, O.~{Nakamura}, N.~A. {Bahcall}, J.~{Brinkmann},
  J.~E. {Gunn}, G.~S. {Hennessy}, {\v Z}.~{Ivezi{\'c}}, G.~R. {Knapp},
  J.~{Loveday}, A.~{Meiksin}, D.~J. {Schlegel}, D.~P. {Schneider},
  I.~{Szapudi}, M.~{Tegmark}, M.~S. {Vogeley}, D.~G. {York}, and {SDSS
  Collaboration}, {\it {The Luminosity and Color Dependence of the Galaxy
  Correlation Function}},  {\em \apj} {\bf 630} (Sept., 2005) 1--27,
  [\href{http://xxx.lanl.gov/abs/astro-ph/}{{\tt astro-ph/}}].

\bibitem{2013MNRAS.430..725V}
F.~C. {van den Bosch}, S.~{More}, M.~{Cacciato}, H.~{Mo}, and X.~{Yang}, {\it
  {Cosmological constraints from a combination of galaxy clustering and lensing
  - I. Theoretical framework}},  {\em \mnras} {\bf 430} (Apr., 2013) 725--746,
  [\href{http://xxx.lanl.gov/abs/1206.6890}{{\tt arXiv:1206.6890}}].

\bibitem{2013MNRAS.430..767C}
M.~{Cacciato}, F.~C. {van den Bosch}, S.~{More}, H.~{Mo}, and X.~{Yang}, {\it
  {Cosmological constraints from a combination of galaxy clustering and lensing
  - III. Application to SDSS data}},  {\em \mnras} {\bf 430} (Apr., 2013)
  767--786, [\href{http://xxx.lanl.gov/abs/1207.0503}{{\tt arXiv:1207.0503}}].

\bibitem{Smith_2009}
R.~E. {Smith}, {\it {Covariance of cross-correlations: towards efficient
  measures for large-scale structure}},  {\em \mnras} {\bf 400} (Dec., 2009)
  851--865, [\href{http://xxx.lanl.gov/abs/0810.1960}{{\tt arXiv:0810.1960}}].

\end{thebibliography}\endgroup

\end{document}